\documentclass{article}
\usepackage{namedplus}
\usepackage{graphicx}
\usepackage{color}
\usepackage{setspace}
\title{\vspace{4cm} Electrostatics in the Stability and Misfolding of the Prion Protein:\\  Salt Bridges, Self-Energy, and Solvation}
\usepackage[in]{fullpage}
\author{Will Guest$^{1}$, Neil Cashman$^{1,3}$, and Steven Plotkin$^{2,3}$}
\date{\today}
\newcommand{\PrPSc}{PrP$^{\textrm{\scriptsize{Sc}}}$}
\newcommand{\PrPC}{PrP$^{\textrm{\scriptsize{C}}}$}
\begin{document}
\maketitle
\begin{center}
\vspace{2cm} Submitted to \textit{Biochemistry and Cell Biology} \\ CSBMCB Special Issue on Protein Misfolding\\
\end{center}
\vspace{6cm}
\noindent$^{1}$Brain Research Centre, University of British Columbia, V6T 2B5\\
$^{2}$Department of Physics and Astronomy, University of British Columbia, V6T 1Z1\\
$^{3}$Corresponding authors (Email:  neil.cashman@vch.ca, steve@physics.ubc.ca, Phone:  1-604-822-2135, 1-604-822-8813)

\newpage
\doublespacing
\section*{Abstract}

Using a recently developed mesoscopic theory of protein dielectrics, we have calculated the salt bridge energies, total residue electrostatic potential energies, and transfer energies into a low dielectric amyloid-like phase for 12 species and mutants of the prion protein.  Salt bridges  and self energies play key roles in stabilizing secondary and tertiary structural elements of the prion protein.  The total electrostatic potential energy of each residue was found to be invariably stabilizing.  Residues frequently found to be mutated in familial prion disease were among those with the largest electrostatic energies.  The large barrier to charged group desolvation imposes regional constraints on involvement of the prion protein in an amyloid aggregate, resulting in an electrostatic amyloid recruitment profile that favours regions of sequence between alpha helix 1 and beta strand 2, the middles of helices 2 and 3, and the region N-terminal to alpha helix 1.  {We found that the stabilization due to salt bridges is minimal among the proteins studied for disease-susceptible human mutants of prion protein.}  

\vspace{2cm}
\noindent\textit{Keywords:}  Salt bridge, prion, protein misfolding, protein electrostatics
\newpage
\section*{Introduction}

Misfolded prion protein is the causative agent for a unique category of human and animal neurodegenerative diseases characterized by progressive dementia, ataxia, and death within months of onset \cite{Prusiner1998}.  These include Creutzfeldt-Jakob disease (CJD), fatal familial insomnia, and Gerstmann-Str\"aussler-Scheinker syndrome in humans, bovine spongiform encephalopathy in cattle, scrapie in sheep, and chronic wasting disease in cervids.  Unlike other infectious conditions that are transmitted by conventional microbes, the material responsible for propagation of prion diseases consists of an abnormally folded conformer of an endogenous protein, possibly in complex with host nucleic acids or sulfated glycans \cite{Caughey2009}.  Soluble, natively-folded monomers of the prion protein (known as \PrPC) may adopt an aggregated protease-resistant conformation known as \PrPSc$\!$ that is capable of recruiting additional monomers of \PrPC$\!$ and inducing them to misfold in a process of template-directed conversion.  This results in ordered multimers of prion protein that, when fractured, act as additional seeds to propagate the misfold through the reservoir of \PrPC$\!$ present in brain.  Although the conversion process may be initiated by an infectious inoculum of \PrPSc, it may also arise spontaneously or due to mutations in the gene coding for PrP that predispose to misfolding.  

Structurally, \PrPC$\!$ is a {glycophosphatidylinositol}-anchored glycoprotein of 232 amino acids comprising an N-terminal unstructured domain and a C-terminal structured domain of 3 $\alpha$-helices (hereafter referred to as $\alpha$1, $\alpha$2, and $\alpha$3 in order) and a short two-stranded antiparallel $\beta$-sheet (made of strands $\beta$1 and $\beta$2), while \PrPSc$\!$ has substantially enriched $\beta$ content speculated to form a stacked $\beta$-helix \cite{Govaerts2004} or extended $\beta$-sheet \cite{Cobb2007} conformation in the amyloid fibril.

At a molecular level PrP misfolding is a physico-chemical process, with the propensity to misfold determined by the free energy difference between folded and misfolded states and the magnitude of the energy barrier separating them.  As in any protein system, electrostatic effects make significant contributions to the energies of the various states and take two forms:  salt bridge energy due to spatial proximity of charged groups within the native protein, and solvation/self energy due to field energy storage in the ambient protein and water dielectric media.  \textit{A priori}, it is expected that electrostatic effects generally favour the well-solvated monomeric \PrPC$\!$ over the more hydrophobic amyloid \PrPSc, since formation of \PrPSc$\!$ necessitates disruption of salt bridges in the native structure (although this may be compensated for by the formation of alternative salt bridges in \PrPSc) and transfer of some charged groups into an environment of lower permittivity, both of which are energetically costly. However, these penalties on formation of \PrPSc$\!$ are counterbalanced by hydrogen bonding, hydrophobic, and possibly entropic contributions that favour the amyloid form \cite{Tsemekhman2007}.  Regional variation in the electrostatic transfer energy to water and amyloid may be useful in predicting participation in the amyloid core of \PrPSc.  Furthermore, several of the causative mutations for familial prion disease involve substitution of charged residues for uncharged residues (such as the D178N mutation responsible for fatal familial insomnia or familial CJD, depending on mutant allele polymorphism status at codon 129) or charge reversal of a residue (such as E200K, the most common mutation in classical familial CJD) \cite{Kovacs2002}, offering an indication of the importance of electrostatic effects in the misfolding process.  {More broadly, it has been found that changes in the charge state of a mutant protein compared to wild-type relate to its tendency to form aggregates \cite{Chiti2003}, and the aggregation propensity of a polypeptide chain is inversely correlated with its net charge \cite{Chiti2002}; similarly, aggregation propensity is maximal at the protein iso-electric point where the net charge is zero \cite{Schmittschmitt2003}.  Intrinsically unstructured proteins tend to have a high net charge \cite{Uversky2000}, which increases the electrostatic cost for the system to condense into the folded structure.}  Sequence correlations between charged groups may affect the kinetics of amyloid formation as well \cite{Dima2004}.

The role of salt bridges in prion disease has been investigated previously by molecular dynamics simulation (MDS) and experimental studies of mutant protein.  MDS of human \PrPC$\!$ has identified salt bridges that play a role in stabilization of the native structure  \cite{Zuegg1999}.  Other MDS studies of the R208H mutation, which disrupts a salt bridge with residues D144 and E146 of $\alpha$-helix 1, {have shown that it} results in global changes to the backbone structure \cite{Bamdad2007}. Experimentally, the E200K mutant of \PrPC$\!$ has been shown through calorimetry to be 4 kJ/mol less stable than wild type \cite{Swietnicki1998}.  Mutation of two aspartates participating in $\alpha$1 intra-helix salt bridges to neutral residues increases misfolding fourfold in cell-free conversion reactions under conditions favouring salt bridge formation \cite{Speare2003}.  Interestingly, complete reversal of charges in $\alpha$1 appears to inhibit conversion, possibly by preventing docking of \PrPC$\!$ and \PrPSc$\!$ \cite{Speare2003}.  The pH dependence of charge interactions in \PrPC$\!$ has also been investigated to identify those most sensitive to pH changes \cite{Warwicker1999}; this is an important aspect of the problem because of the observed increased \PrPC$\!$ misfolding rate at low pH.  

A unifying analysis of all \PrPC$\!$ salt bridges would be useful in understanding their role in structural stability. As well, to our knowledge solvation energy contributions to the misfolding process have not yet been investigated; they would offer a helpful perspective for probing the propensity of different regions of the prion protein to participate in the \PrPSc$\!$ amyloid core.

Direct extraction of salt bridge and solvation energies from molecular dynamics is complicated by the need to run long-length simulations that sample the equilibrium between states of interest, which can be prohibitively slow for states that differ significantly in energy.  An alternative approach is to use a continuum electrostatics description of the protein-water system, in which the response of surrounding material is modelled through solution of the Poisson-Boltzmann equation as a macroscopic dielectric that varies from a low value (usually 4) within the volume of the protein to 78 (the dielectric constant of bulk water) outside the protein.  The downside of this method is that it ignores subtleties of the protein response to perturbing fields, such as cooperative internal reorganization.  Using results from Kirkwood-Fr\"olich theory \cite{OsterG1943,Frohlich1949,VogesD2000}, we have recently developed a procedure to compute a spatially-varying dielectric function for a protein based on fluctuation statistics obtained from brief equilibrium MD simulations that capture much of the microscopic response of the protein at moderate computational cost \cite{Guest2009}.  This provides a convenient tool to calculate solvation and salt bridge energies for all residues in a protein from a single simulation.  In what follows we apply this method to determine the energies for all salt bridges in 12 molecular species of prion protein and the transfer energy for all residues in these proteins into a hypothetical protein amyloid core.

\section*{Methods}

Twelve structures of various species and mutants of \PrPC$\!$ were selected from the Protein Data Bank (PDB), including the species human {1QLZ} and 1QLX \cite{1QLX}, cow {1DX0 }\cite{1DX0}, turtle {1U5L}, frog {1XU0}, chicken {1U3M} \cite{1U5L}, mouse {1AG2} and {1XYX} \cite{1AG2,1XYX}, dog {1XYK}, pig {1XYQ}, cat {1XYJ} \cite{1XYQ}, wallaby {2KFL} \cite{2KFL} and the human mutants D178N {2K1D} \cite{Mills2009} and E200K {1FKC} \cite{1FKC}.  They were taken as starting points for 5ns all-atom molecular dynamics simulations {using the CHARMM force field version C31B1 \cite{CHARMM1983} with explicit pure solvent water (no salt), periodic boundary conditions, particle mesh Ewald electrostatics, a timestep of 2fs, and a Lennard-Jones potential cutoff distance of 13.5\AA.  The basic residues (ARG and LYS) were protonated, while the acidic residues (HIS, ASP, and GLU) were deprotonated to reflect ionization conditions at pH 7.  The system was first minimized for 200 time steps before starting the simulation.}  Snapshots of the simulations were taken every 2ps to build up an ensemble of equilibrium conformations for each protein.  The dipole moments $\mu$ of all residue side chains and backbones were calculated at each snapshot and used to obtain the correlation coefficients
\begin{equation}
R_{ij} = \frac{\left<\mu_i\mu_j\right>}{\sqrt{\left<\mu_i^2\right>\left<\mu_j^2\right>}}
\end{equation}
for all pairs of Cartesian dipole components $\mu_i$ and $\mu_j$, where the angle brackets denote an average over all snapshots (the thermal average).  The matrix R of correlation coefficients was diagonalized to isolate the normal modes of dipole fluctuations, which describe the response of charged groups to perturbations around equilibrium.  The R matrix for each protein was used to calculate the local dielectric map (Guest  et al. 2009).  See Figure 1 in the Supplementary Material for the dielectric map of human PrP.  These dielectric maps were then taken as input for the Poisson-Boltzmann solver APBS \cite{APBS} to solve the linearized Poisson-Boltzmann equation on an 97$^3$ mesh in 150 mM NaCl, again with periodic boundary conditions, to obtain the electrostatic energies required below.  {Atomic radii were assigned according to the CHARMM force field by the program PDB2PQR \cite{Dolinsky2004}.} The often-used simplifying approximation of a constant internal protein dielectric constant of 4 and water dielectric constant of 78 was employed for comparison \cite{Nussinov1999}.

Salt bridges in the set of proteins were identified by searching all pairs of charged atoms for those with charged groups within 12\AA$\!$ of each other, whether the charges were alike or different.  The energy of each salt bridge was determined by a mutation cycle designed to isolate the charge interaction energy from the energy in the surrounding dielectric milieu as shown in Figure 1.  For charged groups A and B, their salt bridge energy $E_{sb}$ was taken to be a function of the energy of the protein system with both charges in place, $E_{AB}$, with one or the other charge removed, $E_A$ and $E_B$, and with both charges removed, $E_0$, as follows:   
\begin{equation}
E_{sb} = E_0 + E_{AB} - \left(E_A + E_B\right).
\end{equation}
Here, $E_0$ contains only the self energy of the part of the protein not including A and B (labelled P in Figure 1), while $E_{AB}$ contains the self energies of A, B, and P  as well as the pairwise interaction energies between A and B, A and P, and B and P.  $E_A$ contains the self energies of A and P and their interaction energies ($E_{B}$ is analogous).  Combining the terms as shown causes all the energies except the interaction energy of A and B to cancel.   

Another cycle, also shown in Figure 1, was used to determine the total contribution of each residue to the electrostatic energy of the protein, $E_{elec}$.  For each side chain in the protein, the electrostatic energies of the side chain $E_{sc}$ and the protein lacking the side chain $E_{whole-sc}$ were calculated in isolation in the protein dielectric environment and subtracted from the electrostatic energy of the intact protein $E_{whole}$: 
\begin{equation}
E_{elec} = E_{whole} - E_{sc} - E_{whole - sc}.
\end{equation} 
The terms $E_{sc}$ and $E_{whole - sc}$ contain the self energies of the side chain and rest of the protein respectively, and $E_{whole}$ contains these self energies as well as the interaction energy of the side chain with the rest of the protein.  Subtracting the terms as shown causes the self energies to cancel, leaving only the interaction energy between the side chain and the protein.  This energy can be thought of as the electrostatic potential energy of a residue in the protein.

To approximate the electrostatic energy of residue transfer into a hydrophobic, low dielectric environment like the core of a \PrPSc$\!$ amyloid, the energy $E(\epsilon_{PrP^{C}})$ of a residue in the dielectric environment of \PrPC$\!$ was compared to the energy $E(\epsilon_{PrP^{Sc}})$ of the residue in a homogeneous dielectric of $\epsilon = 4$, which describes the dielectric response in the interior of a bulk amyloid protein phase. Since the nature of monopole fields in the \PrPSc$\!$ structure is unknown, interactions between charged residues are omitted from the calculation, so the transfer energy reflects only changes in the dielectric environment.  For a given residue, the dielectric contribution to the transfer energy $E_{trans}$ is:
\begin{equation}
E_{trans} = E(\epsilon_{PrP^{Sc}}) - E(\epsilon_{PrP^C}).
\end{equation}

\section*{Results}

\subsection*{Dynamics of Dipoles at Equilibrium}
The modes obtained by diagonalizing the correlation matrix R in Equation 1 generally involved several parts of the molecule; correlations were not limited to residues close in space or sequence.  This is consistent with phonon transmission of perturbations at one site throughout the molecule by strong steric coupling effects through solid-like elastic moduli.  The four largest-amplitude dipole modes for human PrP are shown in Figure 2.  Dipole fluctuations were not qualitatively different between species, but different regions of the molecule exhibited characteristic motions.  The two long alpha helices 2 and 3 exhibited primarily synchronous motion, with the helices rocking back and forth together as a unit.  Nonetheless, some dipoles in the helices exhibited contrary motion.  Alpha helix 1 did show some autonomy from the rest of the structure and tended to fluctuate as a group.
 
Motion of the beta sheet is prominent in several of the modes.  Two patterns stand out:  a see-saw motion in which one strand tilts up as the other tilts down with both strands pivoting about the middle of the strands, and an in-out motion in which the outer beta strand ($\beta$1) and the N-terminal part of $\alpha$2 move synchronously away from the inner beta strand ($\beta$2).  The first motion is seen in modes 2 and 3 in Figure 2 above, while the second motion is seen in other lesser-amplitude modes.  This is compatible with NMR observations of the beta sheet, which show slow exchange between a range of conformations \cite{Liu1999,Viles2001}.  In the NMR experiments, motion of the beta strands was observed on a time scale of microseconds, while these simulations only spanned nanoseconds, but both are indicative of some degree of conformational flexibility in the beta sheet.

\subsection*{Salt bridge energies}

The PrP structures analysed contained a diverse set of salt bridges, ranging from moderately attractive to weakly repulsive.  A complete list of salt bridges in all structures is presented in Table 1 of the Supplementary Material; salt bridges in the human structure are shown in Table 1 {for both the single NMR structure 1QLX and the ensembe of 20 NMR structures 1QLZ}.

Structurally, the salt bridges can be divided into local and nonlocal by the proximity in sequence of the participating residues.  Local salt bridges, like Asp148---Glu152 in $\alpha$1, Asp208---Glu211 in $\alpha$3, and Arg164---Asp167 between $\beta$2 and the following loop serve to stabilize secondary structural elements of the protein, while nonlocal salt bridges like Arg156---Glu196, Arg164---Asp178, and Glu146---Lys204 help to hold these elements together in the overall tertiary fold.  Figure 3 shows the position of these nonlocal salt bridges in bovine PrP.

Many of the salt bridges identified were near the protein surface, where the high degree of solvation attenuates their strength; the strongest salt bridges were those best sequestered from solvent, for this places them in a dielectric environment that increases electric field strength.  The strongest salt bridge of all, between residues 206 and 210 of frog PrP, features a special ``two-pronged'' geometry that enables the amino group of Lys 210 to associate with both carboxyl oxygens on Asp 206.  {Interestingly, two strong but intermittant salt bridges are present in human 1QLZ between the C-terminal arginine and residue 167 in the $\beta2-\alpha2$ loop  and residue 221 in $\alpha3$.  The substantial variation between members of the NMR ensemble at the C-terminus results in large motion of the arginine side chain, so that these salt bridges are only formed in a subset of conformers.  Similarly, the ARG 164 - ASP 178 salt bridge that helps to anchor the beta sheet to $\alpha2$ and $\alpha3$ is not present in all members of the 1QLZ NMR ensemble, although it is quite strong in the single 1QLX structure.}   Although attractive salt bridges predominate, there were a number of repulsive salt bridges identified as seen on the left hand side of Figure 5A, especially in $\alpha$1 and $\alpha$3, which are crowded with several charged residues.  As demonstrated in the following section, despite the presence of these destabilizing interactions, no residue experiences a net repulsive potential as these unfavourable salt bridges are counterbalanced by the presence of other, stronger, favourable ones.  
The total energies due to all salt bridges in each molecular species studied are shown in Figure 5B.  Of note is the much reduced total salt bridge energy in the two human mutants, E200K and D178N, compared to any other structure.  {The categorization of species as susceptible or resistant to prion disease is somewhat approximate, but comparison of total salt bridge energy and disease susceptibility by Kendall's tau gives a value of $\tau =$ 0.45, implying that the order of species by salt bridge energy and disease susceptibility are significantly concordant (p = 0.046).}  Overall, the effect of a heterogeneous dielectric was to moderate putatively strong salt bridges under the biphasic protein-water approximation for the dielectric function. 

The salt bridges listed in Table 1 of the Supplementary Material are those present at pH 7, but for human PrP an additional search was performed to identify salt bridges that would emerge at lower pH, since acidic conditions are known to drive \PrPSc$\!$ formation.  Lower pH results in protonation of histidine residues to produce a positively charged species, which in human PrP enables the formation of three additional weakly attractive salt bridges (indicated by daggers in Table 1).  While the dominant effect of lowering pH is to reprotonate acidic side chains, thus reducing electrostatic stability, this is partially compensated for by the formation of salt bridges involving histidine.

\subsection*{Total residue electrostatic energies}

The salt bridge energies describe pairwise effects, but for mutational analysis it is more important to know the total contribution of each side chain to the stability of the protein.  These energies approximate the electrostatic contribution to the energy change on mutation to a residue with a small nonpolar side chain like alanine. In practice, the side chain of each residue is removed from the protein.  The total electrostatic energy of each residue in all prion proteins studied was less than or equal to 0, indicating a strong degree of evolutionary selection toward residues that benefit stability in the folded conformation.  Through electrically neutral, or nearly so, proteins have their internal dipoles oriented so as to lower the potential energy of every residue.  In human PrP, it is instructive to correlate the energies to known pathogenic mutations:  the residue with the greatest overall stabilizing energy, Thr183, is implicated in familial CJD by a T183A mutation \cite{Kovacs2002}; this mutation has also been shown to radically reduce measured stability by urea denaturation \cite{Liemann1999}.  It is interesting to note that this residue, although not charged, is polar and more deeply buried in the hydrophobic core of the protein than any other charged residue, thereby enhancing the effect of dipolar attractions with its neighbors.  Other residues that on mutation cause familial prion disease have especially high total electrostatic stabilizing energies, including D178 and D202.  Table 3 gives the 10 human side chains with the greatest total electrostatic energies.  We might anticipate that mutation of other residues in Table 3 may enhance the probability of developing misfolding-related disease.

\subsection*{Transfer to hydrophobic environment}

In forming the amyloid core of \PrPSc, some residues must undergo the migration to a region of low dielectric constant. For highly charged residues, this transfer energy is prohibitively high and may thereby exclude their participation in the amyloid core, while for nonpolar residues the small electrostatic transfer energy cost is overcome by favourable solvation entropy changes.  By mapping the transfer energy of each residue into a region of low dielectric approximating \PrPSc$\!$ amyloid, it is possible to predict the likelihood of recruitment for various PrP regions into the amyloid core, without the aid of specific dipole-dipole correlations as might be present in the amyloid.  Figure 6 shows the transfer energy from the \PrPC$\!$ dielectric to a homogeneous dielectric of 4 for various species of \PrPC.  The transfer energy to an aqueous environment would show an inverse pattern.  A 7 amino acid summing window is applied because sequence heterogeneity causes large variation between adjacent residues, and individual residues cannot enter the amyloid core without placing their neighbors in it as well.  The transfer energies in Figure 6 are quite large, but including other terms in addition to the electrostatic energies considered here will reduce the magnitude of the total transfer free energy.

There is considerable variation in the transfer energy along the sequence, with the lowest barrier to dielectric transfer for the region between $\alpha$1 and $\beta$2, the middles of $\alpha$2 and $\alpha$3, and $\beta$1.  Conversely, $\alpha$1, the loop between $\beta$2 and $\alpha$2, and the loop between $\alpha$2 and $\alpha$3 show a formidable barrier to transfer.  This overall pattern is well preserved among all PrP structures studied (see Supplementary Material Figure 4). Immunological studies have defined $\beta$2 as a \PrPSc-specific epitope \cite{Paramithiotis2003}, which presumably necessitates its surface exposure.  In the human structure, $\beta$2 is located at the border between regions of low and high transfer energies, so it is possible that it is close in proximity to the amyloid core but protrudes sufficiently to be recognised by antibodies.  

The overall contour of the transfer energy functions is similar for all PrP structures studied, but there is some variation that correlates with known infectivity data.  As seen in Figure 6, human and bovine share highly similar transfer energy profiles and are both susceptible to prion disease and interspecies transmission of disease, while non-mammalian turtle PrP that does not form \PrPSc$\!$ has a different profile, with a higher transfer energy barrier than cow or human over 4/5 of the sequence.  \PrPC$\!$ from dog, a mammalian species known to be resistant to prion infection \cite{Polymenidou2008}, is intermediate between the human and turtle profiles.  The average transfer energies correlate with a species' resistance to disease (Figure 6 legend). Some other species, including chicken, turtle, and wallaby, have a qualitatively different transfer energy function.

\section*{Discussion}

Although electrostatic effects are only one contribution to the enigmatic \PrPC$\!$ $\rightarrow$ \PrPSc$\!$ conversion process, they offer clues to many of the central questions in prion biochemistry.  The spatial variation of the dielectric is important, as neither charge separation distance, dielectric constant at the midpoint of the salt bridge, nor burial can predict salt bridge stability or total electrostatic energy (see Figures 2 and 3 of the Supplementary Material).  As shown above, salt bridges play an important role in stabilizing both secondary and tertiary aspects of the \PrPC$\!$ structure.  In fact, the total energy of all salt bridges in human \PrPC$\!$ is -60 kJ/mol, almost twice the total stability of the protein as determined by calorimetry \cite{Swietnicki1998,Liemann1999}.  Thus disruption of even a proportion of salt bridges in \PrPC$\!$ is sufficient to substantially destabilize the folded conformation, possibly accelerating or enabling the transition to \PrPSc$\!$ in the right conditions. {However, as has been observed elsewhere \cite{Hendsch1996}, the free energy change of salt bridge disruption may not be equal to the Coulombic energy of the salt bridge itself, due to the competing favourable reduction in desolvation energy.  This will partially offset the change in stability from salt bridge disruption.} 
Charge interactions may also participate in the poorly-understood association between the unstructured N-terminal domain and structured C-terminal domain of \PrPC.  As shown previously, the C-terminal domain organization depends on the length of N-terminal tail present \cite{Li2009}, possibly through a collection of transient interactions below the detection threshold of NMR, resulting in an ``avidity-enhanced'' C-terminal structure.  Charge complementarity between the N- and C-terminal domains provides one explanation for this phenomenon.  For example, the very N-terminus of \PrPC$\!$ contains the highly positively charged region KKRPK from codons 25--29, while $\alpha$1 contains the highly negatively charged region DYED from codons 144--147.  If the N-terminal tail is free to explore a random walk around the C-terminal domain, electrostatic attractions are likely to bias this part of the tail toward residence near $\alpha$1, a region that is especially influenced by the length of tail present.  The net attraction between the N-terminal tail and C-terminal structure may be insufficient to structure the tail but sufficient to collapse or condense the tail onto the surface of the structured domain, resulting in a kind of ``molten shell.''  Further exploration of this phenomenon, by molecular dynamics or other tools, may prove insightful.

The importance of acidic conditions in the PrP conversion has been known for some time \cite{Hornemann1998}, and acidity exerts a large effect through modification of the protonation states of charged residues.  At slightly acidic pH below the pKa of histidine (6.5), protonation of the histidine imidazole ring creates mildly stabilizing salt bridges with nearby residues.  In the prion literature, histidines are a subject of considerable attention for their ability to coordinate copper ions in the octapeptide repeat region of the N-terminal domain \cite{Aronoff2000,Viles2008}, but it seems that they also help to protect \PrPC$\!$ from the stress of mildly acidic conditions.  At much lower pH, however,  protonation of glutamate and aspartate side chains ablates some of the stabilizing salt bridges shown in Table 1, which substantially reduces the energy barrier to rearrangement of \PrPC$\!$ components.  For example, at pH 4.5, the pKa predictor program PROPKA \cite{PROPKA} identifies glutamate residues 168, 200, 219, and 221 as being significantly protonated, which will affect the stability of the protein and whose systematic investigation is a topic for future work.  The influence of acidity on the monomeric \PrPC$\!$ structure has been extensively studied by molecular dynamics \cite{Gu2003,Langella2004,DeMarco2007}, but perhaps the most noteworthy effect of acidity may not be in the resulting structural transition of the isolated \PrPC$\!$ monomer but rather in lowering the barrier to induced reorganization in the presence of the templating species.

Another natural question is the role that electrostatics play in the formation of the \PrPSc$\!$ amyloid.  It may be argued that since the transfer energy profile in Figure 6 neglects the possibility of forming strong salt bridges in the low dielectric amyloid core of \PrPSc$\!$ it misrepresents the ability of these charges to stably occupy the amyloid. A counterexample may be constructed in a case of homogeneous dielectrics.  The total energy change $\Delta E_{total}$ on bringing two opposite charges A and B both of charge $q$ and radius $r_{ion}$ from a large distance apart in a medium with high dielectric $\epsilon_{solv}$ like water into close proximity $r_{AB}$ in a region of low dielectric $\epsilon_{prot}$ to form a salt bridge is equal to the sum of the solvation energy changes $\Delta E_{solv}^A$ and $\Delta E_{solv}^B$ and their pairwise Coulomb energy $\Delta E_{AB}$.  Treating the charges as Born ions, in the limit where $\epsilon_{prot}/\epsilon_{solv} \ll 1$ (a valid assumption since generally $\epsilon_{solv} = 78$ and $\epsilon_{prot} = 4$, so $\epsilon_{prot}/\epsilon_{solv} \approx 0.05$), the total energy change to form the desolvated salt bridge is:
\begin{equation}
\Delta E_{total} = \Delta E_{solv}^A + \Delta E_{solv}^B + \Delta E_{AB} = 2q^2\left(\frac{1}{2r_{ion}}\right)\left(\frac{1}{\epsilon_{prot}} - \frac{1}{\epsilon_{solv}}\right) - \frac{q^2}{\epsilon_{prot}r_{AB}}\approx \frac{q^2}{\epsilon_{prot}}\left(\frac{1}{r_{ion}} - \frac{1}{r_{AB}}\right).
\end{equation}   
This is always positive since $r_{AB}$ is greater than $r_{ion}$ to satisfy the stereochemistry of the atoms.  Thus although salt bridges may partially mitigate the burial of charged residues, they cannot alter the fundamental unfavourability of the electrostatic component of this process.  {It is also possible that solvent-exposed salt bridges may form in the misfolded state outside or on the surface of the amyloid or oligomeric core, which could occur without the desolvation penalty described above.  This would provide a mechanism to stabilize charged and polar parts of the protein in the misfolded form.  Such salt bridges are likely to be relatively low in energy due to the high ambient dielectric environment, and it has been observed for amyloid-beta 16-22 peptide that hydrophobic forces are more important than specific salt bridges in driving amyloid formation \cite{Ma2002}.  However as mentioned above, it has been shown that the net charge of a polypeptide chain incurs resistance to aggregation \cite{Chiti2002,Schmittschmitt2003}, consistent with the notion of a higher overall energetic cost of transfer into a low dielectric medium for more highly charged polypeptides.}  In light of this, we believe the transfer energy profiles accurately convey this part of the obstacle to amyloid formation.  

Continuum electrostatics as a tool to examine protein behaviour has limitations, namely that it ignores the microscopic response of the system and thereby risks omitting subtle but important effects.  However, by deriving the dielectric map from all-atom molecular dynamics simulations of the proteins of interest, we are able to substantially incorporate the microscopic response in our model and thereby improve the reliability of the energy estimates obtained. {Previous theories could not reliably predict the effective dielectric constant inside a protein, so values typically between 4 and 10 have been used as initial guesses.  Stronger salt bridges in the interior tend to be better predicted by an interior dielectric of 4, which would then overestimate the strength of the more abundant salt bridges on the protein surface. An interior dielectric of 10 best predicts the strength of the abundant surface salt bridges, but would then underestimate the strength of the buried interior salt bridges. The heterogeneous dielectric theory in (Guest et al. 2009) makes it unnecessary to guess at the value of the dielectric inside a protein and also indicates that no single value in the interior is satisfactory.}  Quantum effects due to electronic polarizability may be added to this approach as further refinement. {The conformational variability in the ensemble of NMR structures for each PrP molecule also introduces an inherent uncertainty in the calculation of electrostatic energies, which we treated by averaging salt bridge energies over all NMR ensemble members.  The molecular dynamics relaxation methods, often done in the absence of counterions, may introduce uncertainty as well.} Electrostatic considerations are relevant to many aspects of the prion question, from \PrPC$\!$ dynamics and stability to \PrPSc$\!$ amyloid organization and templating.  We have presented an analysis of salt bridge, electrostatic, and hydrophobic transfer energies that provides a useful perspective for understanding the structural vulnerabilities of \PrPC.

\section*{Acknowledgements}

The authors gratefully acknowledge support from the A.P. Sloan Foundation, a donation from William Lambert, the Natural Sciences and Engineering Research Council, PrioNet Canada, the Canadian Institutes for Health Research, the Michael Smith Foundation for Health Research, and the WestGrid computing consortium.

\newpage

\begin{figure}[ht]
\begin{center}
\includegraphics{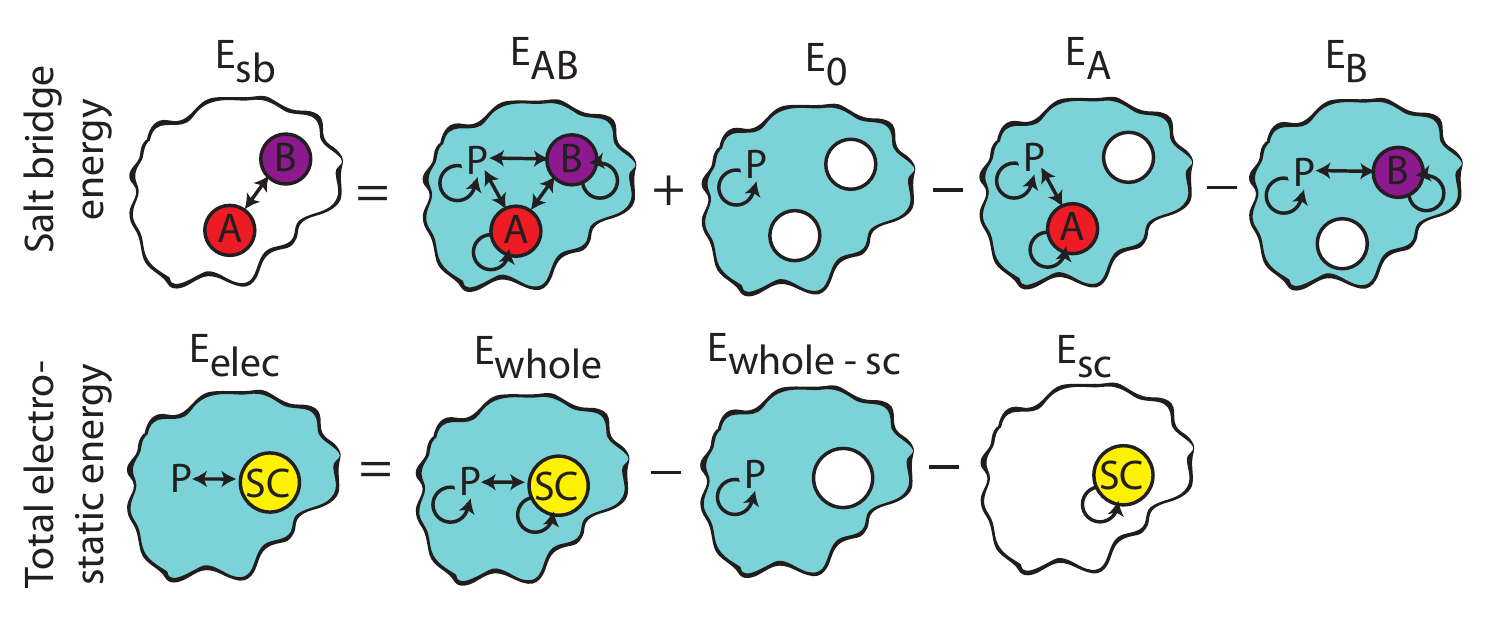}
\caption{Schematic of the approach for calculating salt bridge and total electrostatic energies.  Circular arrows denote solvation/self energies; straight arrows denote interaction energies.  Labels refer to quantities in Equations 2 and 3.}
\end{center}
\end{figure}

\begin{figure}[ht]
\begin{center}
\includegraphics{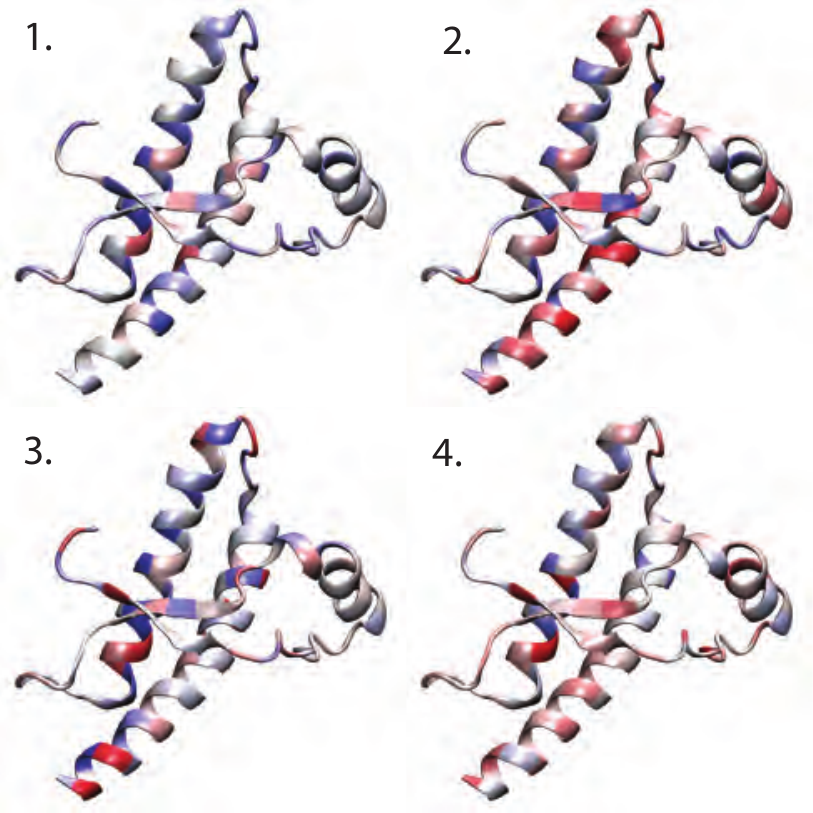}
\caption{The 4 largest amplitude dipole correlation modes for human PrP.  Regions of the same colour move in synchrony, while regions of different colours move in opposition.}
\end{center}
\end{figure}

\begin{figure}[ht]
\begin{center}
\includegraphics{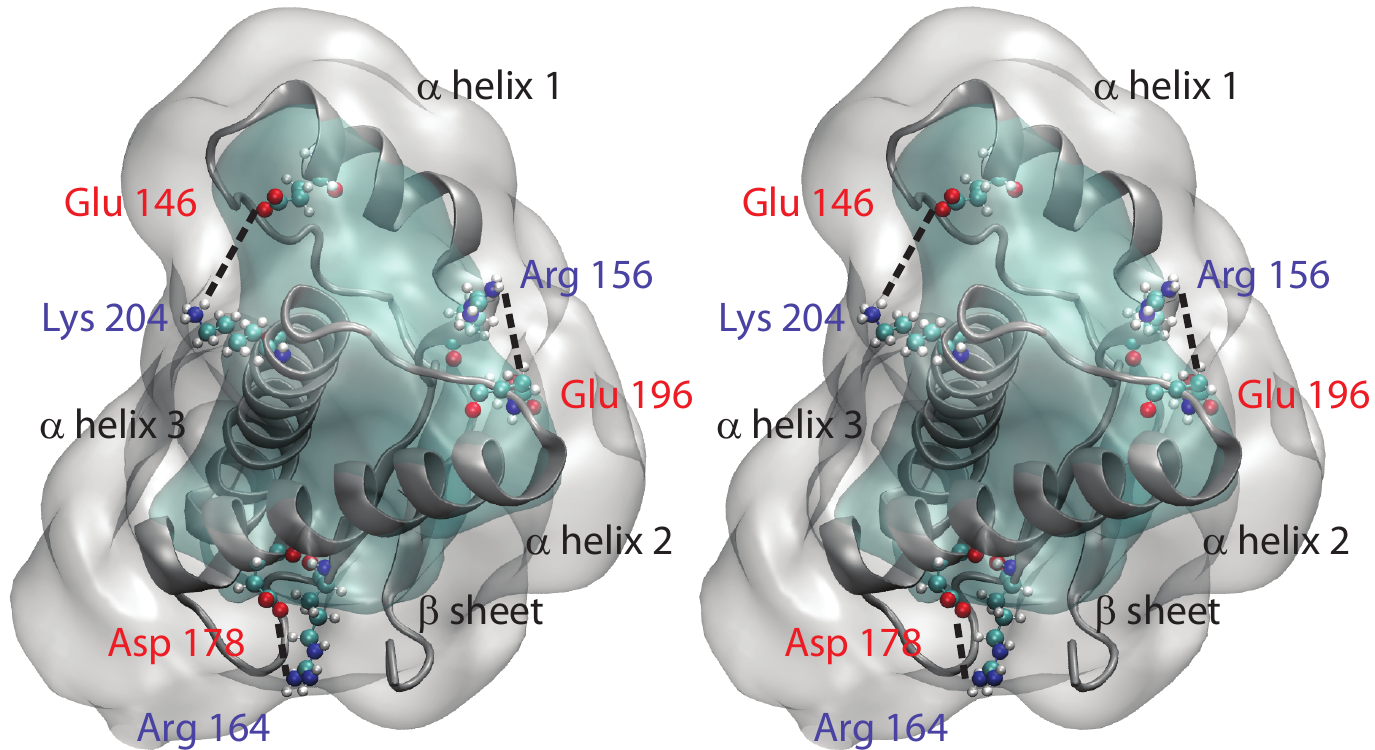}
\caption{{Stereo view of the} three well-conserved nonlocal salt bridges as they are arranged in bovine PrP.  The transparent surfaces show contours of equal dielectric (5 for blue; 70 for white) as determined from heterogeneous mesoscopic dielectric theory (Guest et al. 2009).{The volume near the surface of the protein shows the greatest difference in dielectric on comparison of the homogeneous and heterogeneous dielectric fields.} See also Figure 1 in the supplementary material for a surface plot of the dielectric permittivity as a function of position.}
\end{center}
\end{figure}

\begin{figure}[ht]
\begin{center}
\includegraphics{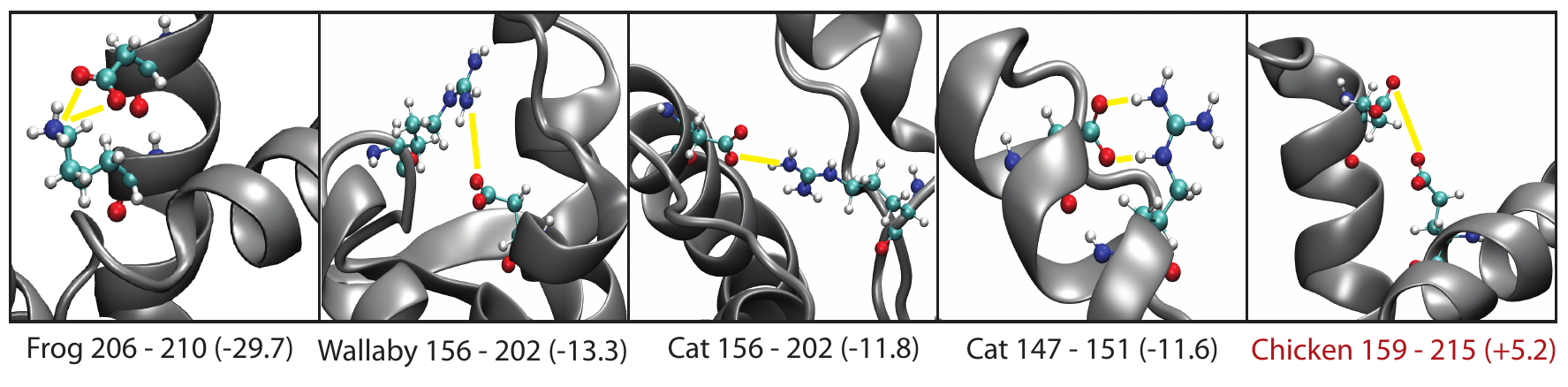}
\caption{The 4 most attractive salt bridges (in black lettering) and the 1 most repulsive (in red lettering) identified in the set of PrP structures. {Numbers in parentheses are the salt bridge energies in kJ/mol for the structures indicated.  Note that interactions with all surrounding residues were considered when assessing the total effect of each residue on overall stability, as given in Table 3 and Supplementary Table 2.}}
\end{center}
\end{figure}

\begin{figure} [ht]
\begin{center}
\includegraphics{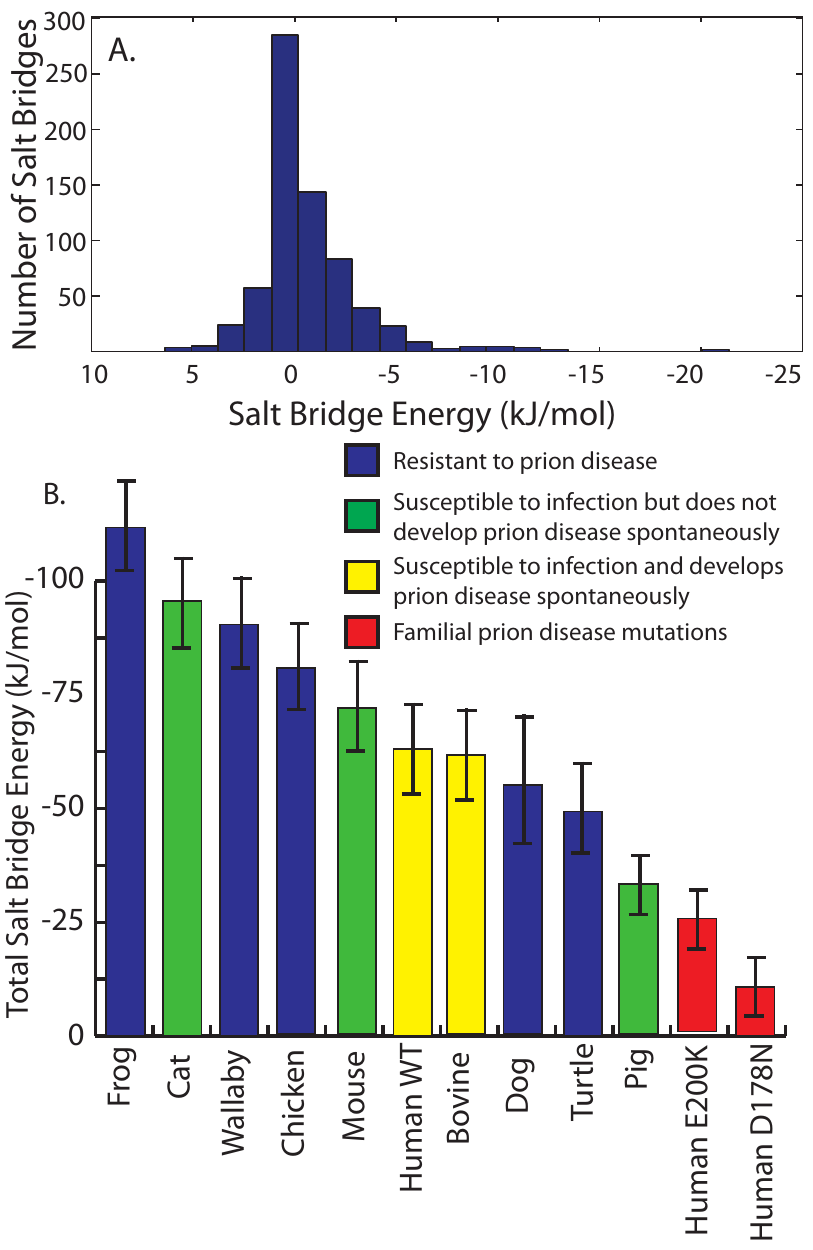}
\caption{A.  Histogram of average energies for all salt bridges identified.  B.  Total salt bridge energies in the molecular species studied.  {Error bars give the 95\% confidence interval for the mean salt bridge energy from each ensemble of NMR structures.}}
\end{center}
\end{figure}

\begin{figure}[ht]
\begin{center}
\includegraphics{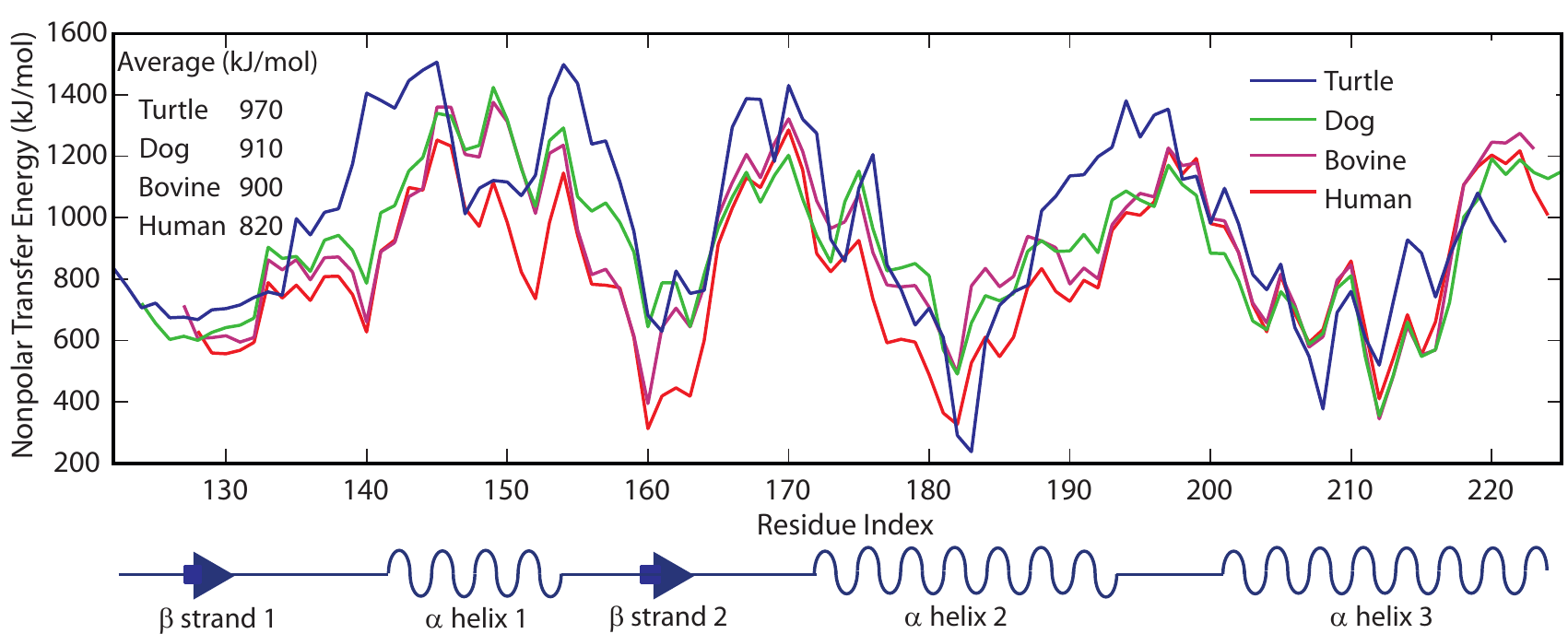}
\caption{Hydrophobic transfer energy for strands of 7 residues centred on a given residue index as calculated from Equation 4 for 4 species of \PrPC. The numbers in the upper left hand corner give the average transfer energy over the whole protein. Shown below are the locations of secondary structural elements in the \PrPC$\!$ sequence.}
\end{center}
\end{figure}

\begin{table}[ht]
\centering
\begin{tabular}{||rlrl|c||c|cc||c|cc||}
\hline
\hline
\multicolumn{4}{||c|}{}& & 1QLX & \multicolumn{2}{|c||}{1QLZ} & 1QLX & \multicolumn{2}{|c||}{1QLZ} \\   
\multicolumn{4}{||c|}{Residues Involved} & r  & $E_{sb}(1)$ & $E_{sb}(1)$ & $\frac{\delta E_{sb}(1)}{E_{sb}(1)}$  & $E_{sb}(2)$ & $E_{sb}(2)$ & $\frac{\delta E_{sb}(2)}{E_{sb}(2)}$  \\
&&&&(\AA)&(kJ/mol)&(kJ/mol) &&(kJ/mol)&(kJ/mol)&\\
\hline
\hline
$\dagger$HIS	&	140	&	ASP	&	147	&	6.6	&	-4.0	&	-3.7	&	0.30	&	-4.6	&	-4.2	&	0.40	\\
ASP	&	144	&	ASP	&	147	&	7.6	&	2.7	&	4.4	&	0.19	&	2.9	&	5.4	&	0.32	\\
GLU	&	146	&	LYS	&	204	&	7.4	&	-2.5	&	-3.0	&	0.29	&	-2.5	&	-3.4	&	0.43	\\
ARG	&	148	&	ARG	&	151	&	7.9	&	1.8	&	2.2	&	1.23	&	1.4	&	2.3	&	0.51	\\
ARG	&	148	&	GLU	&	152	&	4.8	&	-7.4	&	-5.0	&	0.25	&	-35.4	&	-14.6	&	0.51	\\
ARG	&	156	&	GLU	&	196	&	4.9	&	-5.5	&	-5.1	&	0.32	&	-18.2	&	-11.8	&	0.96	\\
ARG	&	156	&	ASP	&	202	&	7.1	&	-3.8	&	-3.0	&	0.27	&	-4.8	&	-3.8	&	0.39	\\
ARG	&	164	&	ASP	&	167	&	6.9	&	-3.0	&	-2.4	&	0.54	&	-3.7	&	-2.7	&	0.95	\\
ARG	&	164	&	ASP	&	178	&	6.0	&	-20.4	&	-4.8	&	1.23	&	-48.2	&	-8.2	&	1.44	\\
$\dagger$HIS	&	177	&	GLU	&	207	&	6.6	&	-2.8	&	-2.8	&	0.43	&	-2.7	&	-3.1	&	0.58	\\
$\dagger$HIS	&	187	&	ASP	&	202	&	8.0	&	-3.0	&	-3.2	&	0.25	&	-5.1	&	-4.6	&	0.30	\\
GLU	&	196	&	ASP	&	202	&	7.9	&	2.9	&	2.4	&	0.22	&	2.9	&	2.7	&	0.25	\\
GLU	&	200	&	LYS	&	204	&	4.1	&	-4.9	&	-3.7	&	0.39	&	-7.5	&	-6.5	&	0.85	\\
ARG	&	208	&	GLU	&	211	&	2.6	&	-9.3	&	-5.8	&	0.30	&	-37.9	&	-14.9	&	0.66	\\
\hline
\end{tabular}
\caption{Salt bridges in the human prion protein from 1QLX {(a single structure) and 1QLZ (an ensemble of 20 structures)}. $\dagger$:  Only present at low pH. The separation between charged groups is given by $r$.  $E_{sb}(1)$ is the salt bridge energy (an average for 1QLZ) calculated with the heterogeneous dielectric theory, and $E_{sb}(2)$ is the same energy calculated with a constant protein dielectric of 4. {The standard deviation of salt bridge energy over the structures in the NMR ensemble containing the salt bridge is given as a fraction of the total salt bridge energy by $\frac{\delta E_{sb}}{E_{sb}}$.  The correlation coefficient between salt bridge energies from 1QLX and 1QLZ is 0.82.}}
\label{1QLXSBs}
\end{table}

\begin{table}[ht]
\centering
\begin{tabular}{|c|c|cccc|c|c|c|c|c|}
\hline
\hline
PDB & Species & \multicolumn{4}{|c|}{Residues Involved} & $n$       & $E_{sb}(1)$ & $\frac{\delta E_{sb}(1)}{E_{sb}(1)}$ & $E_{sb}(2)$ & $\frac{\delta E_{sb}(2)}{E_{sb}(2)}$  \\
        &     &              & & &                      & (/20)    & (kJ/mol) & & (kJ/mol) &       \\ 
\hline
\hline
1XU0	&	Frog	    &ASP&206&LYS&210&	20	&	-21.4	&	0.31	&	-42.2	&	0.37	\\
1QLZ	&	Human WT	&GLU&221&ARG&228&	12	&	-13.7	&	0.07	&	-76.6	&	0.07	\\
2KFL	&	Wallaby	  &ARG&156&ASP&202&	20	&	-13.3	&	0.88	&	-18.0	&	0.90	\\
1QLZ	&	Human WT	&ASP&167&ARG&228&	11	&	-13.1	&	0.06	&	-75.5	&	0.08	\\
1XYJ	&	Cat	      &ARG&156&ASP&202&	20	&	-11.8	&	0.45	&	-21.5	&	0.69	\\
1XYJ	&	Cat	      &ASP&147&ARG&151&	20	&	-11.6	&	0.27	&	-40.9	&	0.33	\\
2KFL	&	Wallaby	  &ARG&156&GLU&196&	20	&	-10.9	&	0.32	&	-29.6	&	0.51	\\
1XYX	&	Mouse    	&ARG&156&ASP&202&	20	&	-10.4	&	0.37	&	-18.9	&	0.46	\\
1QLZ	&	Human WT	&ASP&147&ARG&151&	20	&	-9.9	&	0.38	&	-30.4	&	0.53	\\
1XYX	&	Mouse	    &GLU&146&LYS&204&	20	&	-9.7	&	0.26	&	-27.2	&	0.44	\\
$\vdots$ &$\vdots$ &$\vdots$ &$\vdots$ &$\vdots$ &$\vdots$ &$\vdots$ &$\vdots$ &$\vdots$&$\vdots$ &$\vdots$ \\
2K1D	&	Human D178N	&ARG&156&LYS&194&	20	&	3.4	&	0.42	&	4.9	&	0.84	\\
1XYQ	&	Pig	        &GLU&207&GLU&211&	20	&	3.4	&	0.24	&	4.0	&	0.30	\\
1FKC	&	Human E200K	&LYS&204&ARG&208&	20	&	3.4	&	0.15	&	3.7	&	0.24	\\
1XYQ	&	Pig	        &ARG&148&ARG&151&	20	&	3.5	&	0.09	&	4.1	&	0.24	\\
1XYX	&	Mouse	      &GLU&207&GLU&211&	10	&	3.5	&	0.14	&	4.0	&	0.20	\\
1XYK	&	Dog	        &GLU&207&GLU&211&	20	&	4.0	&	0.18	&	4.8	&	0.25	\\
1QLZ	&	Human WT	  &ASP&144&ASP&147&	19	&	4.4	&	0.18	&	5.4	&	0.31	\\
1XU0	&	Frog	      &LYS&197&LYS&210&	19	&	4.7	&	0.49	&	7.7	&	0.84	\\
2KFL	&	Wallaby	    &GLU&196&ASP&202&	20	&	4.9	&	0.14	&	7.4	&	0.22	\\
1U3M	&	Chicken	    &GLU&159&GLU&215&	20	&	5.2	&	0.15	&	6.2	&	0.43	\\
\hline
\end{tabular}
\caption{The most attractive and repulsive salt bridges in the set of prion protein structures studied. {The number of NMR conformers for each species in which the salt bridge is present is $n$.  $E_{sb}(1)$ is the salt bridge energy (an average for 1QLZ) calculated with the heterogeneous dielectric theory, and $E_{sb}(2)$ is the same energy calculated with a constant protein dielectric of 4.  The standard deviation of salt bridge energy over the structures in the NMR ensemble containing the salt bridge is given as a fraction of the total salt bridge energy by $\frac{\delta E_{sb}}{E_{sb}}$.}}
\end{table}

\begin{table}[ht]
\centering
\begin{tabular}{|c|c|c|c|}
\hline
\hline
Res & Site & $E_{elec}$ (kJ/mol) & $E_{elec}$/ Avg($E_{elec}$)\\
\hline
\hline
\multicolumn{4}{|c|}{1QLX} \\
\hline
\textbf{THR	183}& $\alpha$2	&-197 &  6.0\\
ASP	147& $\alpha$1	&-196         &  5.9\\
TYR	150& $\alpha$1	&-171         &  5.2\\
ARG	136& $\beta$1-$\alpha$1  &-151&  4.6\\
\textbf{ASP	202}& $\alpha$3	&-137 &  4.2\\
ARG	164& $\beta$2 &-99            &  3.0\\
\textbf{ASP	178}& $\alpha$2	&-96  &  2.9\\
GLU	221& $\alpha$3	&-92          &  2.8\\
TYR	157& $\alpha$1-$\beta$2	&-75  &  2.3\\
\textbf{VAL	210}& $\alpha$3 &-73  &  2.2\\
\hline
\multicolumn{4}{|c|}{1QLZ} \\
\hline
\textbf{THR 183} & $\alpha$2  & 	-171 & 5.9 \\
TYR 150	& $\alpha1$           & -171   & 5.9\\
\textbf{ASP 202} & $\alpha$3  &	-135  & 4.6\\
TYR 157 & $\alpha$1-$\beta$2  &	-97   & 3.3\\
ARG 164	& $\beta$2            & -81   & 2.8\\
THR 192	& $\alpha$2-$\alpha$2 & -67   & 2.3\\
\textbf{ASP 178}	& $\alpha2$ & -67   & 2.3\\
\textbf{VAL 210}	& $\alpha3$ & -65   & 2.2\\
ARG 136	& $\beta$1-$\alpha$1  & -63   & 2.1\\
GLU 221	& $\alpha$2           & -61   & 2.1\\
\hline
\end{tabular}
\caption{Residues in human PrP (1QLX and 1QLZ) with the greatest total electrostatic energy from Equation 4.  The last column gives the factor by which each residue's electrostatic energy exceeds the average for all residues in the protein (-33 kJ/mol).  {THR 183 has the lowest electrostatic energy in human PrP and is also a minimum on the hydrophobic transfer profile (see Figure 6), due presumably to its low dielectric local environment.  VAL 210 appears in the list despite being a putatively nonpolar residue because it is located in a region of particularly low dielectric in the protein core, which increases the energy of its small side chain methyl group dipoles.}  $\beta$1-$\alpha$1:  Between $\alpha$1 and $\beta1$. $\alpha1$-$\beta2$:  Between $\alpha1$ and $\beta2$. \textbf{Bold}:  {Wild-type residues at the locations of known mutation sites in familial prion diseases.}  {The correlation between transfer energies from 1QLX and 1QLZ is 0.87. With 90\% confidence, the list is significantly enriched in pathologic mutations compared to random chance (p = 0.096).}}
\end{table}
\end{document}


\maketitle
\begin{figure}[ht]
\begin{center}
\includegraphics{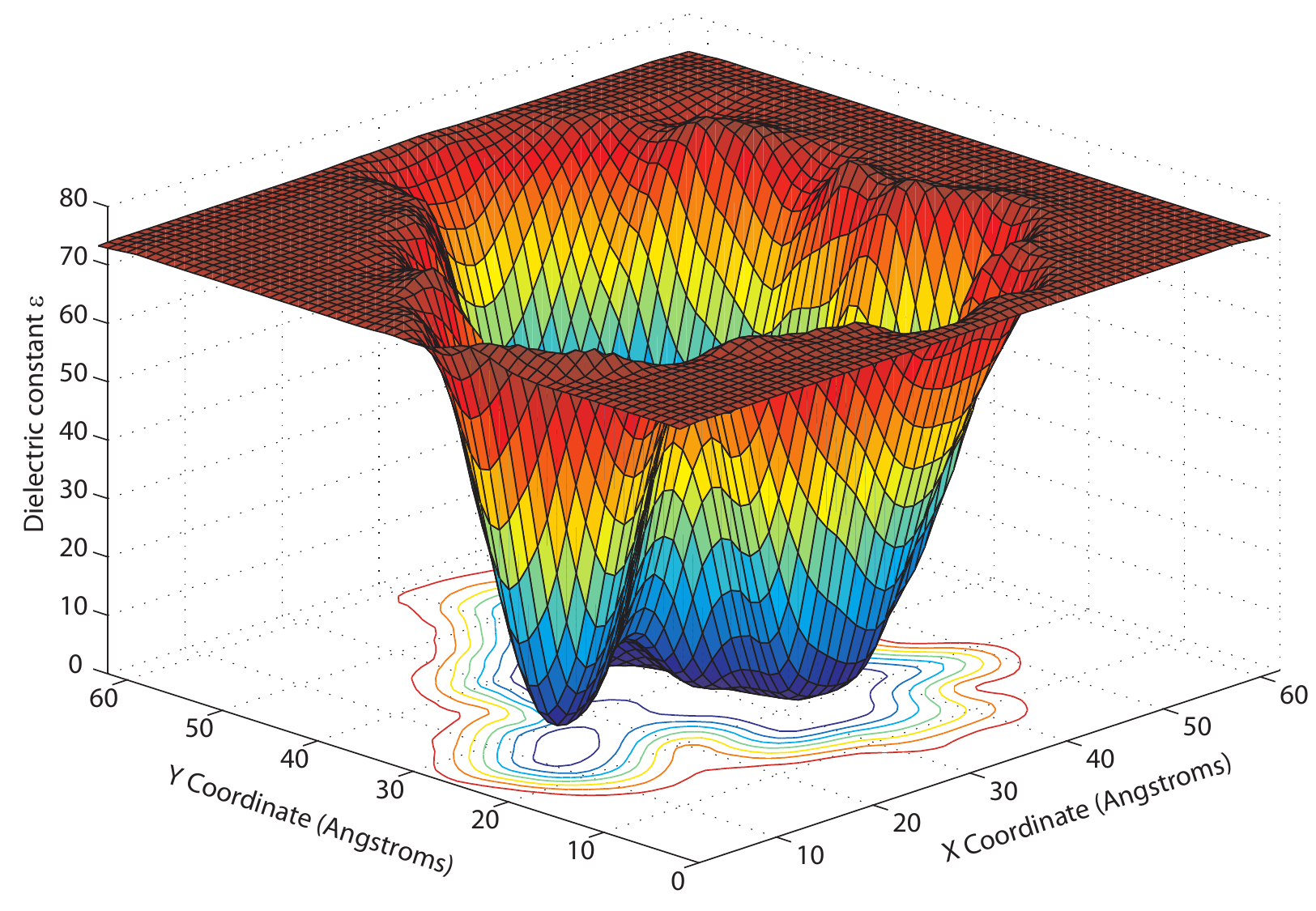}
\caption{Local heterogeneous dielectric map of human PrP as calculated by mesoscopic dielectric theory (Guest et al., 2009).  This is a slice taken through the geometric centre of the molecule in the X-Y plane.}
\end{center}
\end{figure}

\begin{longtable}{|c|c|cc|c|ccc|ccc|}
\caption{Salt bridges in 13 structures of the prion protein with a repulsive or attractive energy greater than 3 kJ/mol.  The number of NMR conformers in which the participating charged groups are within a distance of 12\AA~ is $n$.  $E_{sb}(1)$ is the salt bridge energy calculated according to Equation 2 with the heterogeneous dielectric theory, and $E_{sb}(2)$ is the same energy calculated with a constant protein dielectric of 4.  The standard deviation of salt bridge energy over the structures in the NMR ensemble containing the salt bridge is given by $\delta E_{sb}$ and as a fraction of the total salt bridge energy by $\frac{\delta E_{sb}}{E_{sb}}.$}\\
\hline
\hline
PDB & Species & \multicolumn{2}{|c|}{Residues} & $n$ & $E_{sb}(1)$ & $\delta E_{sb}(1)$ & $\frac{\delta E_{sb}(1)}{E_{sb}(1)}$ & $E_{sb}(2)$ & $\delta E_{sb}(2)$ &$\frac{\delta E_{sb}(2)}{E_{sb}(2)}$  \\
        &     & \multicolumn{2}{|c|}{Involved} & (/20)   & (kJ/mol) & (kJ/mol)  &  & (kJ/mol)& (kJ/mol) &        \\ 
\hline
\hline
\endfirsthead
\hline
\hline
PDB & Species & \multicolumn{2}{|c|}{Residues} & $n$ & $E_{sb}(1)$ & $\delta E_{sb}(1)$&$\frac{\delta E_{sb}(1)}{E_{sb}(1)}$ & $E_{sb}(2)$ & $\delta E_{sb}(2)$&$\frac{\delta E_{sb}(2)}{E_{sb}(2)}$  \\
        &     & \multicolumn{2}{|c|}{Involved} & (/20)   & (kJ/mol) & (kJ/mol) &  & (kJ/mol)& (kJ/mol) &        \\ 
\hline
\hline
\endhead
\multicolumn{9}{|c|}{\textit{Continued$\ldots$}}
\endfoot
\hline
\endlastfoot
1DX0	&	Bovine	&	147	&	151	&	20	&	-6.4	&	3.5	&	0.54	&	-15.2	&	12.6	&	0.83	\\
1DX0	&	Bovine	&	164	&	178	&	20	&	-3.1	&	1.9	&	0.60	&	-6.2	&	6.3	&	1.02	\\
1DX0	&	Bovine	&	156	&	196	&	20	&	-6.0	&	2.8	&	0.47	&	-12.1	&	8.7	&	0.72	\\
1DX0	&	Bovine	&	156	&	202	&	20	&	-7.8	&	5.5	&	0.71	&	-13.0	&	10.7	&	0.82	\\
1DX0	&	Bovine	&	200	&	204	&	20	&	-3.1	&	1.4	&	0.46	&	-4.1	&	5.9	&	1.44	\\
1DX0	&	Bovine	&	208	&	211	&	20	&	-5.1	&	1.9	&	0.38	&	-10.4	&	10.2	&	0.98	\\
	&		&		&		&		&		&		&		&		&		&		\\
1FKC	&	Human E200K	&	164	&	167	&	19	&	-3.0	&	1.1	&	0.38	&	-3.2	&	2.0	&	0.64	\\
1FKC	&	Human E200K	&	164	&	178	&	20	&	-4.6	&	1.0	&	0.23	&	-7.0	&	3.8	&	0.55	\\
1FKC	&	Human E200K	&	156	&	196	&	20	&	-3.1	&	0.7	&	0.22	&	-3.2	&	1.5	&	0.46	\\
1FKC	&	Human E200K	&	156	&	202	&	20	&	-3.2	&	0.7	&	0.21	&	-4.6	&	1.5	&	0.34	\\
1FKC	&	Human E200K	&	146	&	204	&	20	&	-5.2	&	1.6	&	0.31	&	-8.6	&	5.1	&	0.59	\\
1FKC	&	Human E200K	&	204	&	208	&	20	&	3.4	&	0.5	&	0.15	&	3.7	&	0.9	&	0.25	\\
1FKC	&	Human E200K	&	208	&	211	&	20	&	-4.2	&	0.6	&	0.14	&	-5.2	&	1.8	&	0.33	\\
	&		&		&		&		&		&		&		&		&		&		\\
1QLZ	&	Human WT	&	144	&	147	&	19	&	4.4	&	0.8	&	0.19	&	5.4	&	1.7	&	0.32	\\
1QLZ	&	Human WT	&	144	&	151	&	16	&	-3.5	&	1.0	&	0.29	&	-4.1	&	2.5	&	0.60	\\
1QLZ	&	Human WT	&	147	&	151	&	20	&	-9.9	&	3.8	&	0.39	&	-30.4	&	16.2	&	0.53	\\
1QLZ	&	Human WT	&	148	&	152	&	19	&	-5.0	&	1.2	&	0.25	&	-14.6	&	7.5	&	0.51	\\
1QLZ	&	Human WT	&	164	&	168	&	12	&	-3.3	&	1.7	&	0.53	&	-4.9	&	6.5	&	1.32	\\
1QLZ	&	Human WT	&	164	&	178	&	15	&	-4.8	&	5.9	&	1.23	&	-8.2	&	11.8	&	1.44	\\
1QLZ	&	Human WT	&	156	&	196	&	20	&	-5.1	&	1.6	&	0.32	&	-11.8	&	11.4	&	0.96	\\
1QLZ	&	Human WT	&	146	&	204	&	18	&	-3.0	&	0.9	&	0.29	&	-3.4	&	1.4	&	0.43	\\
1QLZ	&	Human WT	&	200	&	204	&	20	&	-3.7	&	1.5	&	0.39	&	-6.5	&	5.5	&	0.85	\\
1QLZ	&	Human WT	&	204	&	207	&	15	&	-3.1	&	1.4	&	0.45	&	-3.7	&	5.9	&	1.58	\\
1QLZ	&	Human WT	&	208	&	211	&	20	&	-5.8	&	1.8	&	0.30	&	-14.9	&	9.9	&	0.66	\\
1QLZ	&	Human WT	&	167	&	228	&	11	&	-13.1	&	0.9	&	0.07	&	-75.5	&	5.7	&	0.07	\\
1QLZ	&	Human WT	&	221	&	228	&	12	&	-13.7	&	1.0	&	0.07	&	-76.6	&	5.7	&	0.07	\\
	&		&		&		&		&		&		&		&		&		&		\\
1U3M	&	Chicken	&	150	&	153	&	20	&	-5.0	&	1.3	&	0.25	&	-10.1	&	7.3	&	0.72	\\
1U3M	&	Chicken	&	155	&	159	&	20	&	-3.8	&	0.7	&	0.19	&	-5.2	&	2.3	&	0.43	\\
1U3M	&	Chicken	&	171	&	185	&	20	&	-5.4	&	1.6	&	0.29	&	-12.4	&	6.7	&	0.54	\\
1U3M	&	Chicken	&	159	&	215	&	20	&	5.2	&	0.8	&	0.15	&	6.2	&	2.7	&	0.44	\\
1U3M	&	Chicken	&	163	&	219	&	20	&	-8.4	&	4.3	&	0.51	&	-20.6	&	15.9	&	0.77	\\
1U3M	&	Chicken	&	217	&	221	&	20	&	-5.7	&	1.1	&	0.20	&	-15.3	&	7.6	&	0.50	\\
1U3M	&	Chicken	&	225	&	229	&	20	&	-4.1	&	1.8	&	0.44	&	-5.9	&	7.7	&	1.30	\\
1U3M	&	Chicken	&	139	&	237	&	20	&	-4.1	&	2.0	&	0.48	&	-6.0	&	6.0	&	0.99	\\
	&		&		&		&		&		&		&		&		&		&		\\
1U5L	&	Turtle	&	143	&	146	&	20	&	-3.3	&	0.7	&	0.21	&	-5.8	&	2.3	&	0.40	\\
1U5L	&	Turtle	&	165	&	169	&	18	&	-3.0	&	2.2	&	0.74	&	-5.0	&	7.1	&	1.42	\\
1U5L	&	Turtle	&	164	&	179	&	20	&	-5.1	&	2.8	&	0.55	&	-7.9	&	7.9	&	1.00	\\
1U5L	&	Turtle	&	175	&	179	&	20	&	-4.2	&	2.8	&	0.68	&	-6.9	&	7.3	&	1.06	\\
1U5L	&	Turtle	&	156	&	202	&	16	&	-3.1	&	2.5	&	0.79	&	-4.2	&	6.8	&	1.63	\\
1U5L	&	Turtle	&	146	&	204	&	20	&	-5.7	&	1.8	&	0.32	&	-15.1	&	10.8	&	0.71	\\
1U5L	&	Turtle	&	191	&	207	&	20	&	-4.6	&	2.6	&	0.56	&	-7.5	&	8.7	&	1.15	\\
	&		&		&		&		&		&		&		&		&		&		\\
1XU0	&	Frog	&	148	&	152	&	20	&	-3.3	&	1.3	&	0.39	&	-4.7	&	6.2	&	1.31	\\
1XU0	&	Frog	&	169	&	176	&	19	&	-3.1	&	1.6	&	0.52	&	-5.6	&	6.6	&	1.17	\\
1XU0	&	Frog	&	175	&	179	&	20	&	-3.1	&	1.3	&	0.43	&	-4.8	&	7.0	&	1.45	\\
1XU0	&	Frog	&	169	&	180	&	19	&	3.0	&	1.4	&	0.46	&	4.1	&	2.8	&	0.69	\\
1XU0	&	Frog	&	188	&	192	&	20	&	-9.4	&	1.8	&	0.19	&	-19.3	&	6.3	&	0.33	\\
1XU0	&	Frog	&	197	&	206	&	20	&	-5.3	&	3.4	&	0.63	&	-10.2	&	11.9	&	1.16	\\
1XU0	&	Frog	&	197	&	210	&	19	&	4.7	&	2.3	&	0.50	&	7.7	&	6.5	&	0.85	\\
1XU0	&	Frog	&	206	&	210	&	20	&	-21.4	&	6.7	&	0.32	&	-42.2	&	15.9	&	0.38	\\
1XU0	&	Frog	&	163	&	221	&	20	&	-9.4	&	6.1	&	0.64	&	-16.9	&	13.6	&	0.80	\\
1XU0	&	Frog	&	221	&	224	&	20	&	-5.0	&	3.6	&	0.72	&	-11.5	&	12.1	&	1.05	\\
	&		&		&		&		&		&		&		&		&		&		\\
1XYJ	&	Cat	&	147	&	151	&	20	&	-11.6	&	3.1	&	0.27	&	-40.9	&	13.4	&	0.33	\\
1XYJ	&	Cat	&	164	&	178	&	20	&	-4.7	&	2.5	&	0.54	&	-11.7	&	11.9	&	1.01	\\
1XYJ	&	Cat	&	156	&	202	&	20	&	-11.8	&	5.3	&	0.44	&	-21.5	&	15.0	&	0.70	\\
1XYJ	&	Cat	&	146	&	204	&	20	&	-3.9	&	2.0	&	0.51	&	-5.7	&	8.7	&	1.51	\\
1XYJ	&	Cat	&	200	&	204	&	20	&	-3.5	&	0.9	&	0.26	&	-4.0	&	3.4	&	0.85	\\
1XYJ	&	Cat	&	208	&	211	&	20	&	-5.8	&	3.0	&	0.52	&	-11.8	&	11.3	&	0.96	\\
	&		&		&		&		&		&		&		&		&		&		\\
1XYK	&	Dog	&	144	&	147	&	20	&	3.1	&	1.0	&	0.32	&	3.9	&	2.0	&	0.52	\\
1XYK	&	Dog	&	144	&	148	&	20	&	-4.1	&	2.1	&	0.51	&	-9.3	&	9.1	&	0.97	\\
1XYK	&	Dog	&	147	&	151	&	20	&	-8.4	&	4.5	&	0.54	&	-22.5	&	16.5	&	0.73	\\
1XYK	&	Dog	&	148	&	152	&	20	&	-3.1	&	1.2	&	0.39	&	-5.3	&	5.8	&	1.09	\\
1XYK	&	Dog	&	164	&	178	&	20	&	-3.6	&	2.2	&	0.60	&	-7.4	&	9.3	&	1.25	\\
1XYK	&	Dog	&	136	&	202	&	2	&	-4.0	&	0.8	&	0.19	&	-6.1	&	0.8	&	0.12	\\
1XYK	&	Dog	&	200	&	204	&	20	&	-3.1	&	1.2	&	0.40	&	-3.8	&	4.7	&	1.22	\\
1XYK	&	Dog	&	204	&	207	&	20	&	-3.8	&	1.4	&	0.38	&	-5.6	&	6.2	&	1.11	\\
1XYK	&	Dog	&	207	&	211	&	20	&	4.0	&	0.7	&	0.16	&	4.8	&	1.2	&	0.26	\\
1XYK	&	Dog	&	208	&	211	&	20	&	-5.3	&	2.5	&	0.48	&	-11.5	&	11.3	&	0.98	\\
	&		&		&		&		&		&		&		&		&		&		\\
1XYQ	&	Pig	&	147	&	151	&	20	&	-4.3	&	1.3	&	0.29	&	-6.0	&	4.1	&	0.69	\\
1XYQ	&	Pig	&	148	&	151	&	20	&	3.5	&	0.3	&	0.09	&	4.1	&	1.0	&	0.24	\\
1XYQ	&	Pig	&	164	&	178	&	20	&	-5.9	&	1.5	&	0.25	&	-11.9	&	7.0	&	0.59	\\
1XYQ	&	Pig	&	200	&	204	&	20	&	-4.0	&	1.3	&	0.33	&	-7.0	&	6.4	&	0.91	\\
1XYQ	&	Pig	&	204	&	207	&	20	&	-3.3	&	1.2	&	0.37	&	-3.0	&	2.1	&	0.71	\\
1XYQ	&	Pig	&	207	&	211	&	20	&	3.4	&	0.8	&	0.23	&	4.0	&	1.2	&	0.30	\\
1XYQ	&	Pig	&	208	&	211	&	20	&	-5.9	&	2.5	&	0.42	&	-9.8	&	12.1	&	1.24	\\
1XYQ	&	Pig	&	220	&	223	&	19	&	-3.8	&	1.2	&	0.33	&	-3.6	&	2.4	&	0.67	\\
	&		&		&		&		&		&		&		&		&		&		\\
1XYX	&	Mouse	&	144	&	147	&	12	&	3.2	&	0.8	&	0.25	&	3.2	&	1.3	&	0.41	\\
1XYX	&	Mouse	&	147	&	151	&	16	&	-5.1	&	2.0	&	0.38	&	-9.2	&	7.9	&	0.85	\\
1XYX	&	Mouse	&	148	&	152	&	12	&	-3.8	&	1.7	&	0.44	&	-7.8	&	9.3	&	1.18	\\
1XYX	&	Mouse	&	164	&	178	&	18	&	-5.4	&	2.4	&	0.45	&	-10.8	&	11.0	&	1.02	\\
1XYX	&	Mouse	&	156	&	196	&	13	&	-3.7	&	1.0	&	0.28	&	-4.2	&	2.5	&	0.61	\\
1XYX	&	Mouse	&	194	&	196	&	16	&	-3.5	&	0.7	&	0.21	&	-3.0	&	0.7	&	0.23	\\
1XYX	&	Mouse	&	156	&	202	&	20	&	-10.4	&	3.9	&	0.37	&	-18.9	&	8.8	&	0.46	\\
1XYX	&	Mouse	&	146	&	204	&	20	&	-9.7	&	2.5	&	0.26	&	-27.2	&	11.9	&	0.44	\\
1XYX	&	Mouse	&	146	&	208	&	10	&	-3.2	&	0.3	&	0.10	&	-3.8	&	0.9	&	0.23	\\
1XYX	&	Mouse	&	204	&	208	&	18	&	3.2	&	0.4	&	0.13	&	3.6	&	0.9	&	0.24	\\
1XYX	&	Mouse	&	207	&	211	&	10	&	3.5	&	0.5	&	0.15	&	4.0	&	0.8	&	0.20	\\
1XYX	&	Mouse	&	208	&	211	&	10	&	-4.4	&	1.1	&	0.24	&	-6.2	&	5.8	&	0.94	\\
1XYX	&	Mouse	&	167	&	221	&	2	&	4.5	&	0.1	&	0.03	&	5.7	&	0.8	&	0.14	\\
1XYX	&	Mouse	&	227	&	230	&	10	&	-3.7	&	1.9	&	0.51	&	-6.9	&	7.9	&	1.15	\\
	&		&		&		&		&		&		&		&		&		&		\\
2K1D	&	Human D178N	&	144	&	147	&	20	&	3.3	&	0.4	&	0.13	&	3.6	&	0.9	&	0.23	\\
2K1D	&	Human D178N	&	147	&	151	&	20	&	-3.7	&	1.1	&	0.30	&	-5.1	&	3.2	&	0.63	\\
2K1D	&	Human D178N	&	164	&	185	&	19	&	3.3	&	1.5	&	0.45	&	3.7	&	2.8	&	0.78	\\
2K1D	&	Human D178N	&	156	&	194	&	20	&	3.4	&	1.4	&	0.40	&	4.9	&	4.1	&	0.85	\\
2K1D	&	Human D178N	&	156	&	202	&	16	&	-3.5	&	1.9	&	0.53	&	-5.5	&	5.1	&	0.93	\\
2K1D	&	Human D178N	&	208	&	211	&	19	&	-3.6	&	1.2	&	0.32	&	-3.0	&	1.4	&	0.46	\\
	&		&		&		&		&		&		&		&		&		&		\\
2KFL	&	Wallaby	&	147	&	151	&	20	&	-5.4	&	2.7	&	0.51	&	-14.6	&	11.3	&	0.77	\\
2KFL	&	Wallaby	&	148	&	152	&	20	&	-3.6	&	1.4	&	0.40	&	-7.9	&	7.6	&	0.96	\\
2KFL	&	Wallaby	&	164	&	178	&	20	&	-4.8	&	1.9	&	0.39	&	-8.3	&	7.8	&	0.93	\\
2KFL	&	Wallaby	&	178	&	185	&	19	&	-3.3	&	0.9	&	0.27	&	-3.7	&	1.6	&	0.42	\\
2KFL	&	Wallaby	&	156	&	196	&	20	&	-10.9	&	3.5	&	0.32	&	-29.6	&	15.1	&	0.51	\\
2KFL	&	Wallaby	&	156	&	202	&	20	&	-13.3	&	11.7	&	0.88	&	-18.0	&	16.2	&	0.90	\\
2KFL	&	Wallaby	&	196	&	202	&	20	&	4.9	&	0.7	&	0.13	&	7.4	&	1.6	&	0.22	\\
2KFL	&	Wallaby	&	200	&	204	&	20	&	-4.6	&	1.1	&	0.24	&	-7.4	&	5.5	&	0.74	\\
2KFL	&	Wallaby	&	204	&	207	&	20	&	-4.2	&	0.7	&	0.17	&	-4.4	&	1.2	&	0.28	\\
2KFL	&	Wallaby	&	146	&	208	&	20	&	-3.7	&	0.3	&	0.09	&	-4.6	&	0.9	&	0.20	\\
2KFL	&	Wallaby	&	208	&	211	&	20	&	-3.8	&	0.9	&	0.25	&	-4.8	&	5.0	&	1.04	\\
\end{longtable}

\begin{figure}[h]
\begin{center}
\includegraphics{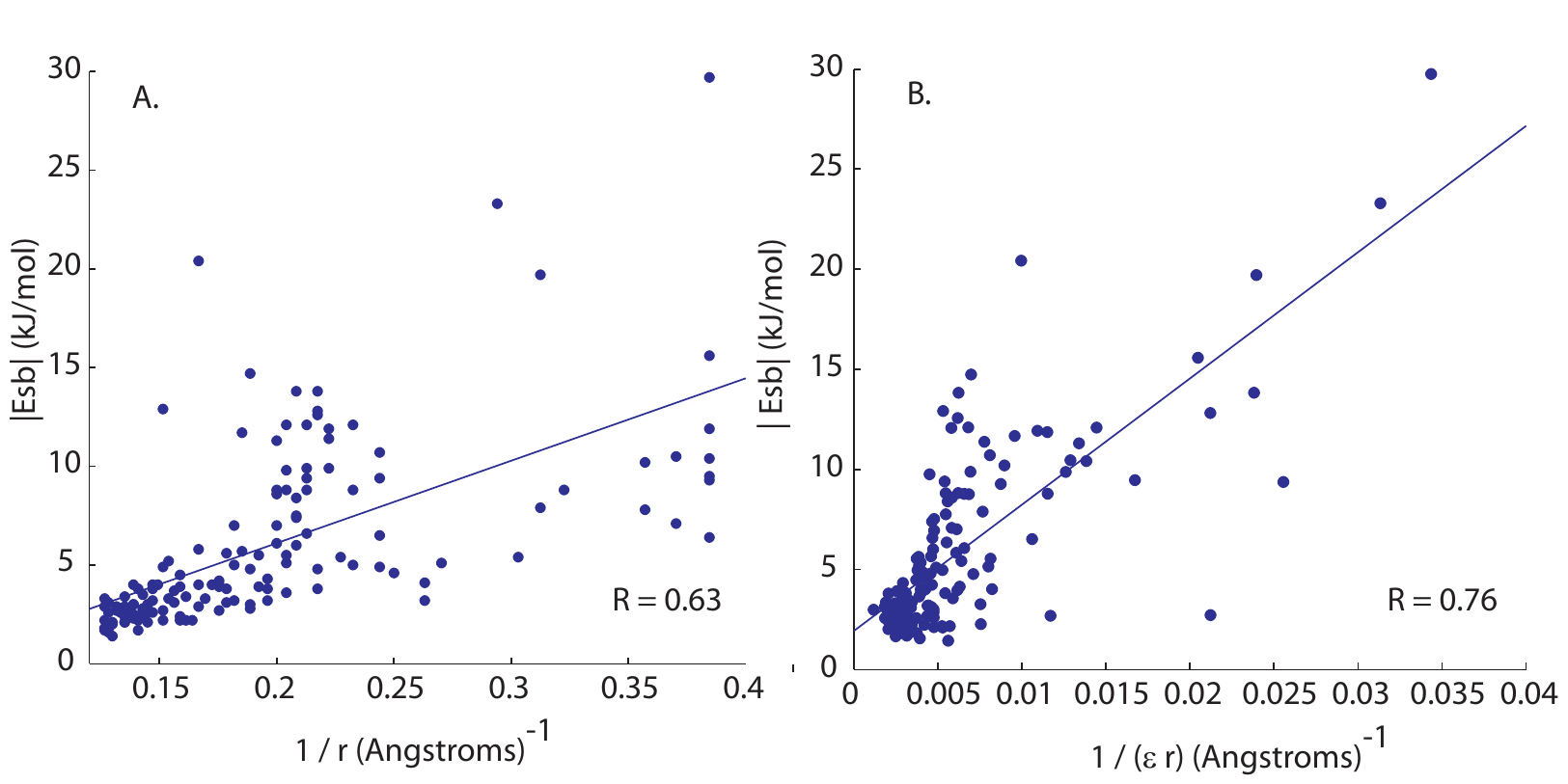}
\caption{A.  Absolute salt bridge energies as a function of the reciprocal of the separation distance between the participating charged groups.  B.  Absolute salt bridge energies as a function of the reciprocal of the separation distance multiplied by the dielectric constant $\epsilon$ at the midpoint of the line joining the charged groups.  In a homogeneous dielectric environment, the energy of a salt bridge connecting charges $q_A$ and $q_B$ would be given by
$
E_{sb} = \frac{1}{4\pi\epsilon_0}\frac{q_Aq_B}{\epsilon \:r}.
$
Including $1/\epsilon$ in the fit in addition to $1/r$ improves the correlation from 0.63 to 0.76, but even taken together the quantities $\epsilon$ and $r$ do not reliably predict salt bridge energies.  Thus a direct calculation of energies from the full heterogeneous dielectric map is necessary to obtain accurate results.}
\end{center}
\end{figure}

\begin{landscape}
\scriptsize
\centering
\begin{longtable}{||c||lr|lr|lr|lr|lr|lr|lr|lr|lr|lr|lr|lr|lr||}
\caption[]{\normalsize{Total electrostatic energies in kJ/mol for every residue in a selection of 13 prion protein structures at pH 7.  The energies are calculated according to Equation 3.}  }\label{MutationEnergies}\\
\hline
\hline
	Res&	\multicolumn{2}{|c}{2KFL}	& \multicolumn{2}{|c}{1QLX}	&	\multicolumn{2}{|c}{1FKC} &	\multicolumn{2}{|c}{2K1D}&	\multicolumn{2}{|c}{1DX0}	&	\multicolumn{2}{|c}{1U3M}	& \multicolumn{2}{|c}{1U5L}	&	\multicolumn{2}{|c}{1XU0}	&	\multicolumn{2}{|c}{1XYJ}	&	\multicolumn{2}{|c}{1XYK}	&	\multicolumn{2}{|c}{1XYQ}	&	\multicolumn{2}{|c}{1XYX}	&	\multicolumn{2}{|c||}{1AG2} \\
	&\multicolumn{2}{|c|}{Wallaby} & \multicolumn{2}{|c}{Human} & \multicolumn{2}{|c}{Human} & \multicolumn{2}{|c}{Human} & \multicolumn{2}{|c}{Bovine} & \multicolumn{2}{|c}{Chicken} & \multicolumn{2}{|c}{Turtle} & \multicolumn{2}{|c}{Frog} & \multicolumn{2}{|c}{Cat} & \multicolumn{2}{|c}{Dog} & \multicolumn{2}{|c}{Pig} & \multicolumn{2}{|c}{Mouse} & \multicolumn{2}{|c||}{Mouse} \\
	&    &    &  \multicolumn{2}{|c}{WT}   & \multicolumn{2}{|c}{E200K} & \multicolumn{2}{|c|}{D178N} &   &     &  &       &  &      &  &    &   &  &   &  &   &  &   &    &  &     \\
\hline
\hline
	&		&		&		&		&		&		&		&		&		&		&		&		&		&		&		&		&		&		&		&		&		&		&		&		&		&		\\
\endfirsthead
\hline
\hline
	Res&	\multicolumn{2}{|c}{2KFL}	& \multicolumn{2}{|c}{1QLX}	&	\multicolumn{2}{|c}{1FKC} &	\multicolumn{2}{|c}{2K1D}&	\multicolumn{2}{|c}{1DX0}	&	\multicolumn{2}{|c}{1U3M}	& \multicolumn{2}{|c}{1U5L}	&	\multicolumn{2}{|c}{1XU0}	&	\multicolumn{2}{|c}{1XYJ}	&	\multicolumn{2}{|c}{1XYK}	&	\multicolumn{2}{|c}{1XYQ}	&	\multicolumn{2}{|c}{1XYX}	&	\multicolumn{2}{|c||}{1AG2} \\
	&\multicolumn{2}{|c|}{Wallaby} & \multicolumn{2}{|c}{Human} & \multicolumn{2}{|c}{Human} & \multicolumn{2}{|c}{Human} & \multicolumn{2}{|c}{Bovine} & \multicolumn{2}{|c}{Chicken} & \multicolumn{2}{|c}{Turtle} & \multicolumn{2}{|c}{Frog} & \multicolumn{2}{|c}{Cat} & \multicolumn{2}{|c}{Dog} & \multicolumn{2}{|c}{Pig} & \multicolumn{2}{|c}{Mouse} & \multicolumn{2}{|c||}{Mouse} \\
	&    &    &  \multicolumn{2}{|c}{WT}   & \multicolumn{2}{|c}{E200K} & \multicolumn{2}{|c|}{D178N} &   &     &  &       &  &      &  &    &   &  &   &  &   &  &   &    &  &     \\
\hline
\hline
	&		&		&		&		&		&		&		&		&		&		&		&		&		&		&		&		&		&		&		&		&		&		&		&		&		&		\\
  \endhead
  \multicolumn{27}{|c|}{\textit{Continued$\ldots$}}
	\endfoot
	\hline
	\hline
	\endlastfoot

120	&	S	&	-11	&		&		&		&		&		&		&		&		&		&		&	S	&	-7	&		&		&		&		&		&		&		&		&		&		&		&		\\
121	&	V	&	-9	&		&		&		&		&		&		&		&		&		&		&	V	&	-2	&		&		&		&		&		&		&		&		&		&		&		&		\\
122	&	V	&	-2	&		&		&		&		&		&		&		&		&		&		&	V	&	-4	&		&		&	V	&	-4	&	V	&	-2	&	V	&	-4	&	V	&	-3	&		&		\\
123	&	G	&	-1	&		&		&		&		&		&		&		&		&		&		&	G	&	0	&		&		&	G	&	-1	&	G	&	-1	&	G	&	-1	&	G	&	-1	&		&		\\
124	&	G	&	-1	&		&		&		&		&		&		&		&		&		&		&	G	&	0	&		&		&	G	&	-1	&	G	&	-1	&	G	&	-1	&	G	&	-1	&		&		\\
125	&	L	&	-19	&		&		&		&		&		&		&	L	&	-7	&		&		&	L	&	-7	&		&		&	L	&	-15	&	L	&	-5	&	L	&	-8	&	L	&	-23	&	L	&	-6	\\
126	&	G	&	-1	&	G	&	-1	&	G	&	-1	&	G	&	-1	&	G	&	-1	&		&		&	G	&	0	&	G	&	-2	&	G	&	-1	&	G	&	-1	&	G	&	-1	&	G	&	-1	&	G	&	-1	\\
127	&	G	&	-1	&	G	&	-1	&	G	&	-1	&	G	&	-2	&	G	&	0	&	S	&	-7	&	G	&	0	&	G	&	-2	&	G	&	0	&	G	&	-1	&	G	&	0	&	G	&	-1	&	G	&	-1	\\
128	&	Y	&	-42	&	Y	&	-32	&	Y	&	-28	&	Y	&	-18	&	Y	&	-52	&	V	&	-2	&	Y	&	-23	&	Y	&	-18	&	Y	&	-21	&	Y	&	-18	&	Y	&	-27	&	Y	&	-28	&	Y	&	-27	\\
129	&	M	&	-11	&	M	&	-9	&	M	&	-8	&	V	&	-6	&	M	&	-7	&	V	&	-2	&	A	&	-2	&	M	&	-8	&	M	&	-7	&	M	&	-5	&	M	&	-4	&	M	&	-9	&	M	&	-8	\\
130	&	L	&	-15	&	L	&	-10	&	L	&	-10	&	L	&	-4	&	L	&	-12	&	G	&	-1	&	L	&	-9	&	L	&	-11	&	L	&	-14	&	L	&	-7	&	L	&	-7	&	L	&	-5	&	L	&	-14	\\
131	&	G	&	-2	&	G	&	-5	&	G	&	-5	&	G	&	-1	&	G	&	-2	&	G	&	0	&	G	&	-2	&	G	&	-3	&	G	&	-2	&	G	&	-1	&	G	&	0	&	G	&	-3	&	G	&	-4	\\
132	&	S	&	-21	&	S	&	-18	&	S	&	-16	&	S	&	-7	&	S	&	-11	&	L	&	-3	&	S	&	-11	&	N	&	-34	&	S	&	-9	&	S	&	-14	&	S	&	-8	&	S	&	-12	&	S	&	-13	\\
133	&	A	&	-1	&	A	&	-3	&	A	&	-1	&	A	&	-1	&	A	&	-1	&	G	&	0	&	A	&	-2	&	A	&	-2	&	A	&	-1	&	A	&	-1	&	A	&	-1	&	A	&	-3	&	A	&	-3	\\
134	&	M	&	-8	&	M	&	-6	&	M	&	-24	&	M	&	-9	&	M	&	-30	&	G	&	0	&	M	&	-10	&	V	&	-19	&	M	&	-11	&	M	&	-6	&	M	&	-8	&	M	&	-8	&	M	&	-22	\\
135	&	S	&	-8	&	S	&	-8	&	S	&	-14	&	S	&	-9	&	S	&	-15	&	Y	&	-10	&	S	&	-16	&	G	&	-1	&	S	&	-38	&	S	&	-19	&	S	&	-10	&	S	&	-9	&	S	&	-11	\\
136	&	R	&	-382	&	R	&	-151	&	R	&	-140	&	R	&	-55	&	R	&	-63	&	A	&	-3	&	G	&	-1	&	R	&	-55	&	R	&	-44	&	R	&	-46	&	R	&	-56	&	R	&	-22	&	R	&	-119	\\
137	&	P	&	-8	&	P	&	-10	&	P	&	-15	&	P	&	-4	&	P	&	-7	&	M	&	-7	&	M	&	-16	&	M	&	-22	&	P	&	-15	&	P	&	-6	&	P	&	-12	&	P	&	-15	&	P	&	-8	\\
138	&	V	&	-6	&	I	&	-11	&	I	&	-10	&	I	&	-5	&	L	&	-7	&	G	&	-3	&	R	&	-35	&	S	&	-15	&	L	&	-4	&	L	&	-22	&	L	&	-5	&	M	&	-3	&	M	&	-4	\\
139	&	M	&	-15	&	I	&	-20	&	I	&	-18	&	I	&	-37	&	I	&	-43	&	R	&	-59	&	M	&	-7	&	Y	&	-16	&	I	&	-18	&	I	&	-9	&	I	&	-20	&	I	&	-16	&	I	&	-26	\\
140	&	H	&	-23	&	H	&	-25	&	H	&	-27	&	H	&	-20	&	H	&	-20	&	V	&	-13	&	N	&	-14	&	Q	&	-13	&	H	&	-12	&	H	&	-15	&	H	&	-25	&	H	&	-16	&	H	&	-15	\\
141	&	F	&	-29	&	F	&	-18	&	F	&	-47	&	F	&	-24	&	F	&	-32	&	M	&	-13	&	F	&	-23	&	F	&	-24	&	F	&	-23	&	F	&	-20	&	F	&	-27	&	F	&	-27	&	F	&	-19	\\
142	&	G	&	1	&	G	&	1	&	G	&	-1	&	G	&	0	&	G	&	0	&	S	&	-16	&	D	&	-26	&	N	&	-15	&	G	&	0	&	G	&	0	&	G	&	0	&	G	&	0	&	G	&	0	\\
143	&	N	&	-17	&	S	&	-9	&	S	&	-19	&	S	&	-10	&	S	&	-26	&	G	&	-1	&	R	&	-32	&	N	&	-27	&	N	&	-33	&	N	&	-14	&	S	&	-10	&	N	&	-12	&	N	&	-15	\\
144	&	E	&	-46	&	D	&	-36	&	D	&	-39	&	D	&	-40	&	D	&	-36	&	M	&	-10	&	P	&	-4	&	P	&	-3	&	D	&	-37	&	D	&	-40	&	D	&	-37	&	D	&	-37	&	D	&	-45	\\
145	&	Y	&	-13	&	Y	&	-11	&	Y	&	-12	&	Y	&	-10	&	Y	&	-11	&	N	&	-11	&	E	&	-29	&	M	&	-6	&	Y	&	-12	&	Y	&	-19	&	Y	&	-10	&	W	&	-19	&	W	&	-18	\\
146	&	E	&	-40	&	E	&	-39	&	E	&	-80	&	E	&	-43	&	E	&	-78	&	Y	&	-40	&	E	&	-90	&	E	&	-127	&	E	&	-54	&	E	&	-70	&	E	&	-43	&	E	&	-51	&	E	&	-45	\\
147	&	D	&	-75	&	D	&	-196	&	D	&	-70	&	D	&	-61	&	D	&	-91	&	H	&	-14	&	R	&	-76	&	S	&	-31	&	D	&	-57	&	D	&	-53	&	D	&	-92	&	D	&	-54	&	D	&	-86	\\
148	&	R	&	-54	&	R	&	-34	&	R	&	-48	&	R	&	-59	&	R	&	-34	&	F	&	-12	&	Q	&	-26	&	R	&	-50	&	R	&	-36	&	R	&	-25	&	R	&	-48	&	R	&	-40	&	R	&	-70	\\
149	&	Y	&	-52	&	Y	&	-20	&	Y	&	-42	&	Y	&	-23	&	Y	&	-27	&	D	&	-32	&	W	&	-57	&	Y	&	-77	&	Y	&	-25	&	Y	&	-31	&	Y	&	-19	&	Y	&	-52	&	Y	&	-106	\\
150	&	Y	&	-86	&	Y	&	-171	&	Y	&	-132	&	Y	&	-68	&	Y	&	-143	&	R	&	-42	&	W	&	-81	&	Y	&	-58	&	Y	&	-49	&	Y	&	-58	&	Y	&	-79	&	Y	&	-52	&	Y	&	-114	\\
151	&	R	&	-97	&	R	&	-43	&	R	&	-55	&	R	&	-43	&	R	&	-69	&	P	&	-2	&	N	&	-47	&	N	&	-57	&	R	&	-51	&	R	&	-37	&	R	&	-61	&	R	&	-36	&	R	&	-46	\\
152	&	E	&	-39	&	E	&	-31	&	E	&	-39	&	E	&	-57	&	E	&	-31	&	D	&	-33	&	E	&	-45	&	D	&	-73	&	E	&	-37	&	E	&	-42	&	E	&	-67	&	E	&	-50	&	E	&	-78	\\
153	&	N	&	-53	&	N	&	-25	&	N	&	-49	&	N	&	-31	&	N	&	-30	&	E	&	-33	&	N	&	-36	&	Y	&	-23	&	N	&	-28	&	N	&	-85	&	N	&	-41	&	N	&	-74	&	N	&	-158	\\
154	&	Q	&	-25	&	M	&	-24	&	M	&	-16	&	M	&	-4	&	M	&	-8	&	Y	&	-17	&	S	&	-10	&	Y	&	-16	&	M	&	-8	&	M	&	-5	&	M	&	-13	&	M	&	-10	&	M	&	-13	\\
155	&	Y	&	-17	&	H	&	-19	&	H	&	-27	&	H	&	-19	&	H	&	-18	&	R	&	-41	&	N	&	-15	&	N	&	-13	&	Y	&	-13	&	Y	&	-9	&	Y	&	-13	&	Y	&	-13	&	Y	&	-21	\\
156	&	R	&	-128	&	R	&	-29	&	R	&	-41	&	R	&	-60	&	R	&	-97	&	W	&	-36	&	R	&	-77	&	Q	&	-14	&	R	&	-53	&	R	&	-68	&	R	&	-57	&	R	&	-133	&	R	&	-242	\\
157	&	Y	&	-145	&	Y	&	-75	&	Y	&	-148	&	Y	&	-18	&	Y	&	-140	&	W	&	-52	&	Y	&	-9	&	M	&	-41	&	Y	&	-109	&	Y	&	-66	&	Y	&	-52	&	Y	&	-128	&	Y	&	-228	\\
158	&	P	&	-16	&	P	&	-12	&	P	&	-9	&	P	&	-6	&	P	&	-37	&	S	&	-13	&	P	&	-3	&	P	&	-32	&	P	&	-9	&	P	&	-4	&	P	&	-5	&	P	&	-10	&	P	&	-12	\\
159	&	N	&	-21	&	N	&	-32	&	N	&	-52	&	N	&	-14	&	N	&	-38	&	E	&	-54	&	N	&	-59	&	N	&	-17	&	N	&	-17	&	D	&	-54	&	N	&	-16	&	N	&	-17	&	N	&	-22	\\
160	&	Q	&	-23	&	Q	&	-34	&	Q	&	-18	&	Q	&	-24	&	Q	&	-16	&	N	&	-56	&	Q	&	-27	&	R	&	-80	&	Q	&	-26	&	Q	&	-23	&	Q	&	-34	&	Q	&	-26	&	Q	&	-24	\\
161	&	V	&	-63	&	V	&	-53	&	V	&	-64	&	V	&	-26	&	V	&	-54	&	S	&	-20	&	V	&	-34	&	V	&	-61	&	V	&	-50	&	V	&	-27	&	V	&	-43	&	V	&	-51	&	V	&	-49	\\
162	&	M	&	-19	&	Y	&	-25	&	Y	&	-25	&	Y	&	-34	&	Y	&	-32	&	A	&	-1	&	Y	&	-29	&	Y	&	-38	&	Y	&	-32	&	Y	&	-34	&	Y	&	-21	&	Y	&	-40	&	Y	&	-50	\\
163	&	Y	&	-59	&	Y	&	-51	&	Y	&	-40	&	Y	&	-31	&	Y	&	-39	&	R	&	-30	&	Y	&	-42	&	R	&	-83	&	Y	&	-45	&	Y	&	-26	&	Y	&	-23	&	Y	&	-36	&	Y	&	-30	\\
164	&	R	&	-59	&	R	&	-99	&	R	&	-63	&	R	&	-62	&	R	&	-74	&	Y	&	-15	&	K	&	-30	&	P	&	-28	&	R	&	-52	&	R	&	-82	&	R	&	-84	&	R	&	-56	&	R	&	-182	\\
165	&	P	&	-16	&	P	&	-5	&	P	&	-5	&	P	&	-8	&	P	&	-5	&	P	&	-5	&	E	&	-47	&	M	&	-7	&	P	&	-3	&	P	&	-3	&	P	&	-3	&	P	&	-5	&	P	&	-4	\\
166	&	I	&	-25	&	M	&	-19	&	M	&	-21	&	M	&	-2	&	V	&	-13	&	N	&	-48	&	Y	&	-152	&	Y	&	-33	&	V	&	-15	&	V	&	-7	&	V	&	-24	&	V	&	-16	&	V	&	-34	\\
167	&	D	&	-34	&	D	&	-41	&	D	&	-58	&	D	&	-48	&	D	&	-31	&	R	&	-67	&	N	&	-12	&	R	&	-35	&	D	&	-35	&	D	&	-38	&	D	&	-54	&	D	&	-32	&	D	&	-39	\\
168	&	Q	&	-20	&	E	&	-31	&	E	&	-39	&	E	&	-70	&	Q	&	-11	&	V	&	-47	&	D	&	-34	&	G	&	-1	&	Q	&	-18	&	Q	&	-16	&	Q	&	-17	&	Q	&	-15	&	Q	&	-20	\\
169	&	Y	&	-137	&	Y	&	-16	&	Y	&	-8	&	Y	&	-7	&	Y	&	-8	&	Y	&	-31	&	R	&	-48	&	E	&	-93	&	Y	&	-7	&	Y	&	-6	&	Y	&	-15	&	Y	&	-40	&	Y	&	-10	\\
170	&	G	&	-1	&	S	&	-25	&	S	&	-27	&	S	&	-9	&	S	&	-18	&	Y	&	-52	&	S	&	-10	&	E	&	-41	&	S	&	-28	&	S	&	-25	&	S	&	-8	&	S	&	-13	&	S	&	-10	\\
171	&	S	&	-12	&	N	&	-20	&	N	&	-17	&	N	&	-13	&	N	&	-19	&	R	&	-36	&	V	&	-22	&	Y	&	-10	&	N	&	-15	&	N	&	-28	&	N	&	-18	&	N	&	-13	&	N	&	-13	\\
172	&	Q	&	-22	&	Q	&	-11	&	Q	&	-21	&	Q	&	-15	&	Q	&	-30	&	D	&	-29	&	P	&	-3	&	V	&	-17	&	Q	&	-15	&	Q	&	-15	&	Q	&	-24	&	Q	&	-12	&	Q	&	-70	\\
173	&	N	&	-15	&	N	&	-11	&	N	&	-16	&	N	&	-13	&	N	&	-18	&	Y	&	-131	&	E	&	-36	&	S	&	-14	&	N	&	-10	&	N	&	-13	&	N	&	-17	&	N	&	-13	&	N	&	-17	\\
174	&	S	&	-14	&	N	&	-12	&	N	&	-14	&	N	&	-17	&	N	&	-15	&	S	&	-7	&	G	&	-1	&	E	&	-49	&	N	&	-11	&	N	&	-11	&	S	&	-11	&	N	&	-13	&	N	&	-34	\\
175	&	F	&	-79	&	F	&	-53	&	F	&	-54	&	F	&	-28	&	F	&	-45	&	S	&	-12	&	R	&	-134	&	D	&	-36	&	F	&	-30	&	F	&	-18	&	F	&	-73	&	F	&	-24	&	F	&	-79	\\
176	&	V	&	-18	&	V	&	-9	&	V	&	-19	&	V	&	-19	&	V	&	-14	&	P	&	-3	&	F	&	-68	&	R	&	-47	&	V	&	-6	&	V	&	-14	&	V	&	-46	&	V	&	-10	&	V	&	-10	\\
177	&	H	&	-17	&	H	&	-17	&	H	&	-16	&	H	&	-16	&	H	&	-18	&	V	&	-4	&	V	&	-9	&	F	&	-56	&	H	&	-14	&	R	&	-29	&	H	&	-15	&	H	&	-11	&	H	&	-17	\\
178	&	D	&	-69	&	D	&	-96	&	D	&	-63	&	N	&	-32	&	D	&	-56	&	P	&	-3	&	R	&	-39	&	V	&	-10	&	D	&	-42	&	D	&	-52	&	D	&	-39	&	D	&	-41	&	D	&	-78	\\
179	&	C	&	-14	&	C	&	-15	&	C	&	-12	&	C	&	-11	&	C	&	-13	&	Q	&	-24	&	D	&	-127	&	R	&	-46	&	C	&	-6	&	C	&	-8	&	C	&	-11	&	C	&	-9	&	C	&	-16	\\
180	&	V	&	-22	&	V	&	-11	&	V	&	-12	&	V	&	-18	&	V	&	-9	&	D	&	-31	&	C	&	-13	&	D	&	-60	&	V	&	-16	&	V	&	-15	&	V	&	-38	&	V	&	-9	&	V	&	-6	\\
181	&	N	&	-19	&	N	&	-14	&	N	&	-18	&	N	&	-17	&	N	&	-14	&	V	&	-3	&	V	&	-21	&	C	&	-11	&	N	&	-13	&	N	&	-20	&	N	&	-24	&	N	&	-9	&	N	&	-19	\\
182	&	I	&	-30	&	I	&	-29	&	I	&	-14	&	I	&	-31	&	I	&	-18	&	F	&	-60	&	N	&	-19	&	Y	&	-13	&	I	&	-26	&	I	&	-9	&	I	&	-26	&	I	&	-10	&	I	&	-31	\\
183	&	T	&	-212	&	T	&	-197	&	T	&	-172	&	T	&	-158	&	T	&	-193	&	V	&	-20	&	I	&	-36	&	N	&	-18	&	T	&	-166	&	T	&	-157	&	T	&	-165	&	T	&	-174	&	T	&	-219	\\
184	&	V	&	-14	&	I	&	-34	&	I	&	-37	&	I	&	-22	&	V	&	-14	&	A	&	-1	&	T	&	-178	&	M	&	-15	&	V	&	-41	&	V	&	-8	&	V	&	-48	&	I	&	-12	&	I	&	-10	\\
185	&	K	&	-21	&	K	&	-19	&	K	&	-23	&	K	&	-29	&	K	&	-23	&	D	&	-37	&	V	&	-31	&	S	&	-133	&	R	&	-50	&	K	&	-16	&	K	&	-21	&	K	&	-18	&	K	&	-18	\\
186	&	Q	&	-51	&	Q	&	-23	&	Q	&	-26	&	Q	&	-20	&	E	&	-88	&	C	&	-13	&	T	&	-19	&	V	&	-9	&	Q	&	-20	&	Q	&	-58	&	Q	&	-24	&	Q	&	-28	&	Q	&	-33	\\
187	&	H	&	-29	&	H	&	-37	&	H	&	-43	&	H	&	-34	&	H	&	-41	&	F	&	-22	&	E	&	-59	&	T	&	-26	&	H	&	-26	&	H	&	-76	&	H	&	-19	&	H	&	-50	&	H	&	-39	\\
188	&	T	&	-29	&	T	&	-22	&	T	&	-34	&	T	&	-56	&	T	&	-34	&	N	&	-15	&	Y	&	-20	&	E	&	-84	&	T	&	-66	&	T	&	-11	&	T	&	-56	&	T	&	-18	&	T	&	-24	\\
189	&	T	&	-21	&	V	&	-4	&	V	&	-6	&	V	&	-4	&	V	&	-4	&	I	&	-14	&	K	&	-13	&	Y	&	-107	&	V	&	-7	&	V	&	-3	&	V	&	-8	&	V	&	-3	&	V	&	-4	\\
190	&	T	&	-16	&	T	&	-18	&	T	&	-11	&	T	&	-12	&	T	&	-26	&	T	&	-133	&	I	&	-29	&	I	&	-13	&	T	&	-18	&	T	&	-17	&	T	&	-26	&	T	&	-15	&	T	&	-28	\\
191	&	T	&	-19	&	T	&	-23	&	T	&	-19	&	T	&	-73	&	T	&	-99	&	V	&	-20	&	D	&	-149	&	I	&	-8	&	T	&	-22	&	T	&	-31	&	T	&	-51	&	T	&	-54	&	T	&	-26	\\
192	&	T	&	-49	&	T	&	-51	&	T	&	-50	&	T	&	-33	&	T	&	-16	&	T	&	-18	&	P	&	-4	&	K	&	-21	&	T	&	-27	&	T	&	-8	&	T	&	-20	&	T	&	-19	&	T	&	-26	\\
193	&	T	&	-11	&	T	&	-17	&	T	&	-21	&	T	&	-8	&	T	&	-16	&	E	&	-43	&	N	&	-15	&	P	&	-4	&	T	&	-13	&	T	&	-8	&	T	&	-13	&	T	&	-10	&	T	&	-15	\\
194	&	K	&	-16	&	K	&	-19	&	K	&	-15	&	K	&	-16	&	K	&	-20	&	Y	&	-43	&	E	&	-50	&	A	&	-2	&	K	&	-14	&	K	&	-25	&	K	&	-19	&	K	&	-18	&	K	&	-16	\\
195	&	G	&	0	&	G	&	0	&	G	&	0	&	G	&	0	&	G	&	0	&	S	&	-8	&	N	&	-36	&	E	&	-29	&	G	&	0	&	G	&	0	&	G	&	0	&	G	&	0	&	G	&	-1	\\
196	&	E	&	-53	&	E	&	-30	&	E	&	-41	&	E	&	-37	&	E	&	-48	&	I	&	-14	&	Q	&	-13	&	G	&	0	&	E	&	-83	&	E	&	-67	&	E	&	-46	&	E	&	-40	&	E	&	-78	\\
197	&	N	&	-22	&	N	&	-12	&	N	&	-18	&	N	&	-18	&	N	&	-13	&	G	&	-1	&	N	&	-13	&	K	&	-19	&	N	&	-11	&	N	&	-13	&	N	&	-11	&	N	&	-19	&	N	&	-18	\\
198	&	F	&	-34	&	F	&	-10	&	F	&	-34	&	F	&	-11	&	F	&	-42	&	P	&	-2	&	V	&	-19	&	N	&	-17	&	F	&	-24	&	F	&	-21	&	F	&	-16	&	F	&	-35	&	F	&	-23	\\
199	&	T	&	-37	&	T	&	-29	&	T	&	-55	&	T	&	-32	&	T	&	-16	&	A	&	-2	&	T	&	-14	&	N	&	-22	&	T	&	-17	&	T	&	-34	&	T	&	-16	&	T	&	-30	&	T	&	-63	\\
200	&	E	&	-35	&	E	&	-37	&	K	&	-17	&	E	&	-33	&	E	&	-30	&	A	&	-1	&	Q	&	-17	&	S	&	-8	&	E	&	-31	&	E	&	-31	&	E	&	-33	&	E	&	-36	&	E	&	-33	\\
201	&	T	&	-34	&	T	&	-20	&	T	&	-25	&	T	&	-20	&	T	&	-14	&	K	&	-9	&	V	&	-6	&	E	&	-27	&	T	&	-14	&	T	&	-25	&	T	&	-13	&	T	&	-23	&	T	&	-25	\\
202	&	D	&	-304	&	D	&	-137	&	D	&	-290	&	D	&	-122	&	D	&	-235	&	K	&	-14	&	E	&	-104	&	L	&	-14	&	D	&	-123	&	D	&	-340	&	D	&	-86	&	D	&	-161	&	D	&	-372	\\
203	&	I	&	-9	&	V	&	-5	&	V	&	-5	&	V	&	-8	&	I	&	-7	&	N	&	-14	&	V	&	-18	&	N	&	-16	&	M	&	-8	&	M	&	-5	&	V	&	-7	&	V	&	-6	&	V	&	-7	\\
204	&	K	&	-12	&	K	&	-9	&	K	&	-12	&	K	&	-8	&	K	&	-7	&	T	&	-12	&	R	&	-46	&	Q	&	-15	&	K	&	-15	&	K	&	-11	&	K	&	-13	&	K	&	-12	&	K	&	-11	\\
205	&	I	&	-58	&	M	&	-20	&	M	&	-40	&	M	&	-13	&	M	&	-42	&	S	&	-10	&	V	&	-74	&	L	&	-14	&	I	&	-39	&	I	&	-70	&	M	&	-15	&	M	&	-40	&	M	&	-18	\\
206	&	M	&	-49	&	M	&	-37	&	M	&	-46	&	M	&	-17	&	M	&	-39	&	E	&	-27	&	M	&	-28	&	D	&	-119	&	M	&	-24	&	M	&	-34	&	I	&	-42	&	M	&	-47	&	M	&	-32	\\
207	&	E	&	-61	&	E	&	-44	&	E	&	-36	&	E	&	-60	&	E	&	-42	&	A	&	-2	&	K	&	-23	&	T	&	-14	&	E	&	-51	&	E	&	-36	&	E	&	-69	&	E	&	-39	&	E	&	-45	\\
208	&	R	&	-35	&	R	&	-35	&	R	&	-32	&	R	&	-73	&	R	&	-41	&	V	&	-4	&	Q	&	-16	&	T	&	-13	&	R	&	-58	&	R	&	-30	&	R	&	-45	&	R	&	-32	&	R	&	-27	\\
209	&	V	&	-86	&	V	&	-66	&	V	&	-58	&	V	&	-30	&	V	&	-56	&	A	&	-1	&	V	&	-60	&	V	&	-51	&	V	&	-50	&	V	&	-41	&	V	&	-84	&	V	&	-47	&	V	&	-63	\\
210	&	V	&	-80	&	V	&	-73	&	V	&	-73	&	V	&	-55	&	V	&	-75	&	A	&	-3	&	I	&	-81	&	K	&	-95	&	V	&	-67	&	V	&	-61	&	V	&	-82	&	V	&	-66	&	V	&	-86	\\
211	&	E	&	-52	&	E	&	-41	&	E	&	-42	&	E	&	-59	&	E	&	-44	&	A	&	-2	&	Q	&	-16	&	S	&	-11	&	E	&	-53	&	E	&	-47	&	E	&	-94	&	E	&	-37	&	E	&	-56	\\
212	&	Q	&	-24	&	Q	&	-17	&	Q	&	-21	&	Q	&	-104	&	Q	&	-20	&	N	&	-18	&	E	&	-46	&	Q	&	-19	&	Q	&	-33	&	Q	&	-23	&	Q	&	-25	&	Q	&	-14	&	Q	&	-23	\\
213	&	M	&	-31	&	M	&	-27	&	M	&	-38	&	M	&	-30	&	M	&	-41	&	Q	&	-13	&	M	&	-31	&	I	&	-82	&	M	&	-36	&	M	&	-26	&	M	&	-35	&	M	&	-36	&	M	&	-38	\\
214	&	C	&	-14	&	C	&	-12	&	C	&	-16	&	C	&	-13	&	C	&	-14	&	T	&	-12	&	C	&	-12	&	I	&	-54	&	C	&	-8	&	C	&	-12	&	C	&	-11	&	C	&	-7	&	C	&	-19	\\
215	&	I	&	-7	&	I	&	-7	&	I	&	-17	&	I	&	-28	&	I	&	-10	&	E	&	-44	&	M	&	-5	&	R	&	-40	&	V	&	-10	&	V	&	-7	&	I	&	-24	&	V	&	-5	&	V	&	-33	\\
216	&	T	&	-17	&	T	&	-24	&	T	&	-57	&	T	&	-51	&	T	&	-20	&	V	&	-9	&	Q	&	-29	&	E	&	-50	&	T	&	-23	&	T	&	-24	&	T	&	-30	&	T	&	-28	&	T	&	-50	\\
217	&	Q	&	-66	&	Q	&	-53	&	Q	&	-82	&	Q	&	-21	&	Q	&	-54	&	E	&	-25	&	Q	&	-36	&	M	&	-52	&	Q	&	-43	&	Q	&	-53	&	Q	&	-60	&	Q	&	-45	&	Q	&	-46	\\
218	&	Y	&	-34	&	Y	&	-23	&	Y	&	-51	&	Y	&	-91	&	Y	&	-22	&	M	&	-3	&	Y	&	-22	&	C	&	-14	&	Y	&	-40	&	Y	&	-37	&	Y	&	-48	&	Y	&	-22	&	Y	&	-122	\\
219	&	Q	&	-17	&	E	&	-37	&	E	&	-44	&	E	&	-26	&	Q	&	-15	&	E	&	-55	&	Q	&	-28	&	I	&	-10	&	Q	&	-19	&	Q	&	-17	&	Q	&	-27	&	Q	&	-15	&	Q	&	-32	\\
220	&	N	&	-25	&	R	&	-53	&	R	&	-77	&	R	&	-33	&	R	&	-56	&	N	&	-13	&	Q	&	-18	&	T	&	-23	&	K	&	-16	&	K	&	-15	&	K	&	-16	&	K	&	-20	&	K	&	-16	\\
221	&	E	&	-220	&	E	&	-92	&	E	&	-77	&	E	&	-88	&	E	&	-91	&	K	&	-9	&	Y	&	-33	&	E	&	-196	&	E	&	-42	&	E	&	-43	&	E	&	-58	&	E	&	-53	&	E	&	-43	\\
222	&	Y	&	-18	&	S	&	-19	&	S	&	-40	&	S	&	-14	&	S	&	-11	&	V	&	-29	&	Q	&	-56	&	Y	&	-64	&	S	&	-52	&	S	&	-25	&	Y	&	-16	&	S	&	-29	&	S	&	-43	\\
223	&	Q	&	-22	&	Q	&	-15	&	Q	&	-17	&	Q	&	-13	&	Q	&	-17	&	V	&	-19	&	L	&	-8	&	R	&	-41	&	E	&	-39	&	E	&	-43	&	E	&	-79	&	Q	&	-13	&	Q	&	-16	\\
224	&	A	&	-2	&	A	&	-2	&	A	&	-2	&	A	&	-2	&	A	&	-3	&	T	&	-24	&	A	&	-2	&	R	&	-63	&	A	&	-2	&	A	&	-1	&	A	&	-2	&	A	&	-2	&	A	&	-2	\\
225	&	A	&	-2	&	Y	&	-16	&	Y	&	-26	&	Y	&	-17	&	Y	&	-21	&	K	&	-15	&		&		&	G	&	0	&	Y	&	-11	&	Y	&	-11	&	Y	&	-11	&	Y	&	-11	&	Y	&	-15	\\
226	&	Q	&	-21	&	Y	&	-9	&	Y	&	-22	&	Y	&	-9	&	Y	&	-13	&	V	&	-16	&		&		&		&		&	Y	&	-35	&	Y	&	-11	&	A	&	-1	&	Y	&	-11	&		&		\\
227	&	R	&	-46	&	Q	&	-14	&	Q	&	-24	&	Q	&	-14	&		&		&	I	&	-56	&		&		&		&		&	Q	&	-18	&	Q	&	-22	&	Q	&	-13	&	D	&	-29	&		&		\\
228	&	Y	&	-9	&		&		&	R	&	-41	&	R	&	-30	&		&		&	R	&	-47	&		&		&		&		&	R	&	-28	&	R	&	-33	&	R	&	-37	&	G	&	0	&		&		\\
229	&	Y	&	-11	&		&		&	G	&	0	&	G	&	0	&		&		&	E	&	-47	&		&		&		&		&	R	&	-29	&	G	&	1	&	G	&	-1	&	R	&	-39	&		&		\\
230	&		&		&		&		&	S	&	-8	&	S	&	-9	&		&		&	M	&	-41	&		&		&		&		&	A	&	-1	&	A	&	-1	&	A	&	-1	&	R	&	-26	&		&		\\
231	&		&		&		&		&		&		&		&		&		&		&	C	&	-9	&		&		&		&		&		&		&		&		&		&		&	S	&	-8	&		&		\\
232	&		&		&		&		&		&		&		&		&		&		&	V	&	-12	&		&		&		&		&		&		&		&		&		&		&		&		&		&		\\
233	&		&		&		&		&		&		&		&		&		&		&	Q	&	-22	&		&		&		&		&		&		&		&		&		&		&		&		&		&		\\
234	&		&		&		&		&		&		&		&		&		&		&	Q	&	-67	&		&		&		&		&		&		&		&		&		&		&		&		&		&		\\
235	&		&		&		&		&		&		&		&		&		&		&	Y	&	-41	&		&		&		&		&		&		&		&		&		&		&		&		&		&		\\
236	&		&		&		&		&		&		&		&		&		&		&	R	&	-35	&		&		&		&		&		&		&		&		&		&		&		&		&		&		\\
237	&		&		&		&		&		&		&		&		&		&		&	E	&	-55	&		&		&		&		&		&		&		&		&		&		&		&		&		&		\\
238	&		&		&		&		&		&		&		&		&		&		&	Y	&	-22	&		&		&		&		&		&		&		&		&		&		&		&		&		&		\\
239	&		&		&		&		&		&		&		&		&		&		&	R	&	-36	&		&		&		&		&		&		&		&		&		&		&		&		&		&		\\
240	&		&		&		&		&		&		&		&		&		&		&	L	&	-3	&		&		&		&		&		&		&		&		&		&		&		&		&		&		\\
241	&		&		&		&		&		&		&		&		&		&		&	A	&	-1	&		&		&		&		&		&		&		&		&		&		&		&		&		&		\\
\end{longtable}
\end{landscape}

\normalsize
\begin{figure}[h]
\begin{center}
\includegraphics{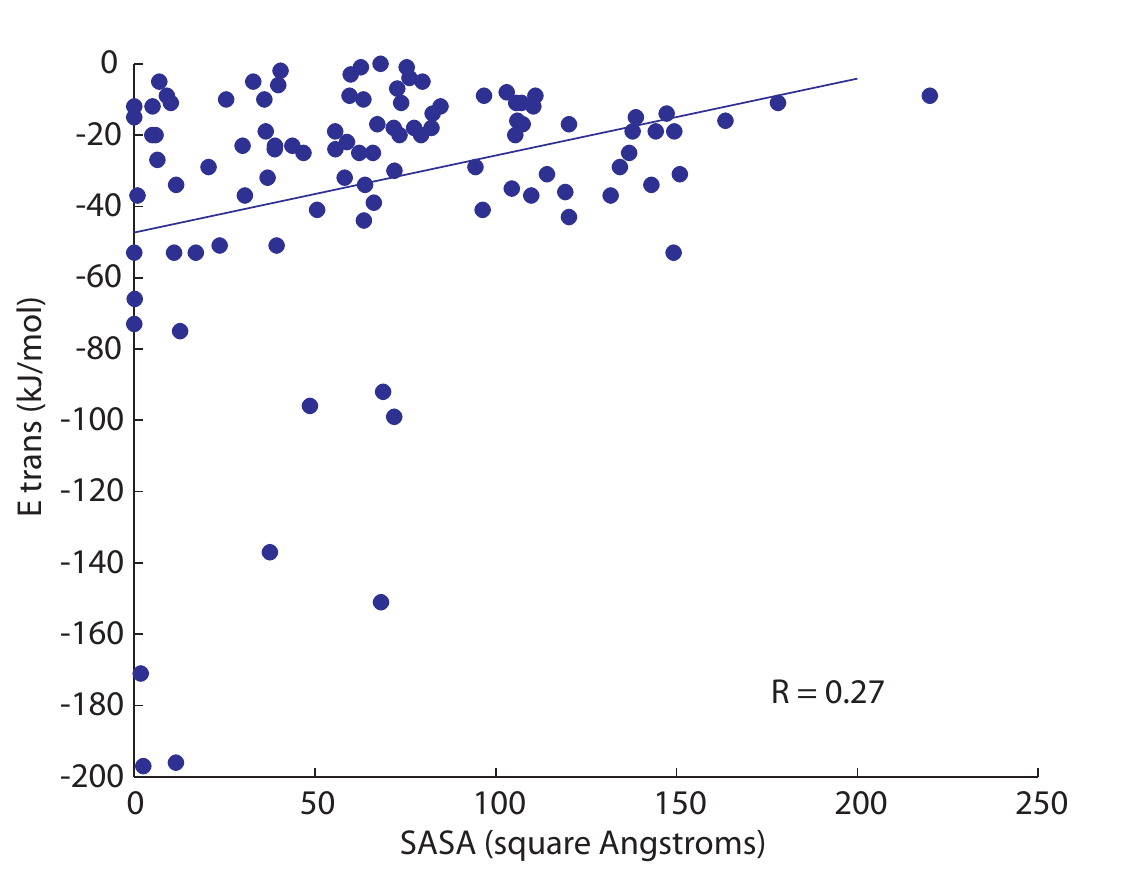}
\caption{Total electrostatic energies of side chains in human PrP calculated from Equation 3 plotted as a function of solvent-accessible surface area (SASA), a statistic related to burial in the core of the protein.  There is a weak but significant (p = 0.01) correlation between the energy and SASA, but SASA alone is a poor proxy for the electrostatic energy.}
\end{center}
\end{figure}

\begin{figure}[ht]
\begin{center}
\includegraphics{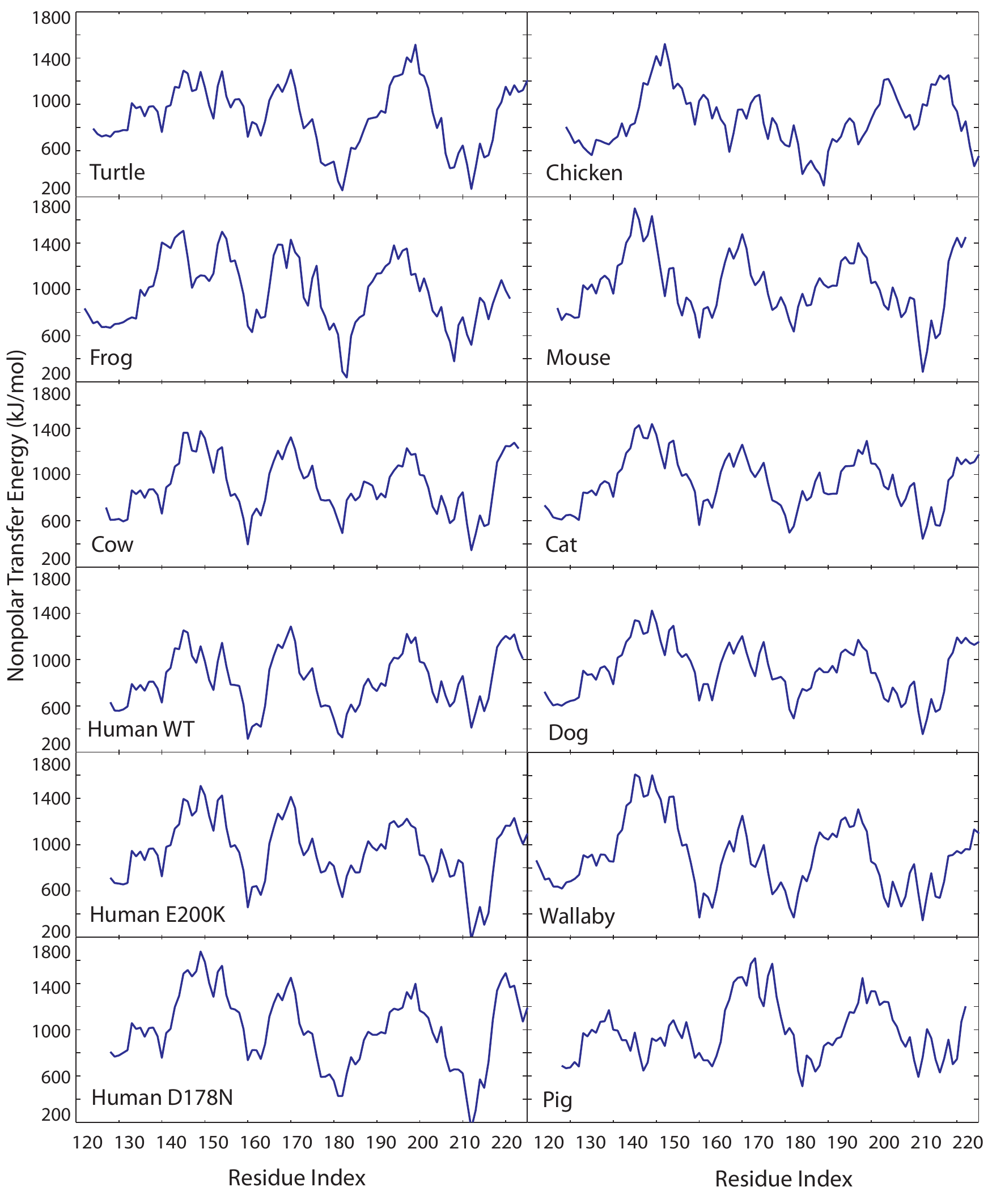}
\caption{Hydrophobic transfer energies from Equation 4 for seven residue contiguous segments of 12 prion protein structures.  The residue index gives the position of the middle residue in each segment. }
\end{center}
\end{figure}